\documentclass[journal]{IEEEtran}

\ifCLASSINFOpdf
\else
\fi
\usepackage[figuresright]{rotating}
\usepackage{graphicx, subcaption, color, cite, epsfig, dsfont, amssymb, amsmath, amsthm, amsfonts, amsbsy, mathrsfs, mathtools, delarray}

\newtheorem{theorem}{Theorem}

\newtheorem{lemma}{Lemma}
\newtheorem{algorithm}{Algorithm}
\newtheorem{remark}{Remark}

\begin{document}

%
\title{Off-policy Reinforcement Learning for $ H_\infty $ Control Design}
%
%
%
\author{Biao~Luo,
        Huai-Ning~Wu
        and~Tingwen~Huang
\thanks{Manuscript received November 24, 2013; revised March 28, 2014 and April 2, 2014; accepted April 18, 2014. This work was supported in part by the National Basic Research Program of China under 973 Program of Grant 2012CB720003, in part by the National Natural Science Foundation of China under Grant 61121003, and in part by the General Research Fund project from Science and Technology on Aircraft Control Laboratory of Beihang University under Grant 9140C480301130C48001. This publication was also made possi- ble by NPRP grant \# NPRP 4-1162-1-181 from the Qatar National Research Fund (a member of Qatar Foundation). The statements made herein are solely the responsibility of the authors. This paper was recommended by Associate Editor H. M. Schwartz.}
\thanks{Biao Luo is with the Science and Technology on Aircraft Control Laboratory, Beihang University (Beijing University of Aeronautics and Astronautics), Beijing 100191, P. R. China. He is also with State Key Laboratory of Management and Control for Complex Systems, Institute of Automation, Chinese Academy of Sciences, Beijing 100190, P. R. China (E-mail: biao.luo@hotmail.com).}
\thanks{Huai-Ning Wu is with the Science and Technology on Aircraft Control Laboratory, Beihang University (Beijing University of Aeronautics and Astronautics), Beijing 100191, P. R. China (E-mail:whn@buaa.edu.cn).}
\thanks{Tingwen~Huang is with the Texas A\& M University at Qatar, PO Box 23874, Doha, Qatar (E-mail: tingwen.huang@qatar.tamu.edu).}
\thanks{Digital Object Identifier 10.1109/TCYB.2014.2319577}}

%
%

\markboth{IEEE Transactions on Cybernetics,~Vol. 99~, No. PP, In Press, 2014}%
{Shell \MakeLowercase{\textit{et al.}}: Bare Demo of IEEEtran.cls for Journals}
%



\maketitle

\begin{abstract}
The $H_\infty$ control design problem is considered for nonlinear systems with unknown internal system model. It is known that the nonlinear $ H_\infty $ control problem can be transformed into solving the so-called Hamilton-Jacobi-Isaacs (HJI) equation, which is a nonlinear partial differential equation that is generally impossible to be solved analytically. Even worse, model-based approaches cannot be used for approximately solving HJI equation, when the accurate system model is unavailable or costly to obtain in practice. To overcome these difficulties, an off-policy reinforcement leaning (RL)  method is introduced to learn the solution of HJI equation from real system data instead of mathematical system model, and its convergence is proved. In the off-policy RL method, the system data can be generated with arbitrary policies rather than the evaluating policy, which is extremely important and promising for practical systems. For implementation purpose, a neural network (NN) based actor-critic structure is employed and a least-square NN weight update algorithm is derived based on the method of weighted residuals. Finally, the developed NN-based off-policy RL method is tested on a linear F16 aircraft plant, and further applied to a rotational/translational actuator system. 

\end{abstract}


%
\IEEEpeerreviewmaketitle

\section{Introduction}
%
%
%
%

\IEEEPARstart{R}{einforcement}  learning (RL) is a machine learning technique that has been widely studied from the computational intelligence and machine learning scope in the artificial intelligence community  \cite{kaelbling1996reinforcement, sutton1998reinforcement, bertsekas2005dynamic, powell2007approximate}. RL technique refers to an actor or agent that interacts with its environment and aims to learn the optimal actions, or control policies, by observing their responses from the environment. In  \cite{sutton1998reinforcement}, Sutton and Barto suggested a definition of RL method, i.e., any method that is well suited for solving RL problem can be regarded as a RL method, where the RL problem is defined in terms of optimal control of Markov decision processes. This obviously established the relationship between the RL method and control community. Moreover, RL methods have the ability to find an optimal control policy from unknown environment, which makes RL a promising method for control design of real systems. 

In the past few years, many RL approaches \cite{murray2002adaptive, yadav2007robust,  zhang2008novel, vrabie2009adaptive,vrabie2009neural, zhang2009neural, zhang2011data, luo2012online, vrabie2012optimal, liu2012neural,  luo2012approximate, wu2012heuristic, liu2013finite, wei2013novel, liu2014decentralized, luo2013data, modares2014integral, liu2014policy, luo2014databased} have been introduced for solving the optimal control problems. Especially, some extremely important results were reported by using RL for solving the optimal control problem of discrete-time systems \cite{zhang2008novel, zhang2009neural, liu2012neural, liu2013finite, wei2013novel, liu2014policy}.
Such as, 
Liu and Wei suggested a finite-approximation-error based iterative adaptive dynamic programming approach \cite{liu2013finite}, and a novel policy iteration (PI) method \cite{liu2014policy} for discrete-time nonlinear systems.
For continuous-time systems, Murray et al. \cite{murray2002adaptive} presented two PI algorithms that avoid the necessity of knowing the internal system dynamics.
Vrabie et al. \cite{vrabie2009adaptive, vrabie2009neural, vrabie2012optimal} introduced the thought of PI and proposed an important framework of integral reinforcement learning (IRL).
Modares et al. \cite{modares2014integral} developed an experience replay based IRL algorithm for nonlinear partially unknown constrained-input systems. In \cite{liu2014decentralized}, an online neural network (NN) based decentralized control strategy was developed for stabilizing a class of continuous-time nonlinear interconnected large-scale systems. In addition, it worth mentioning that the thought of RL methods have also been introduced to solve the optimal control problem of partial differential equation systems \cite{yadav2007robust, luo2012online, luo2012approximate, wu2012heuristic, luo2014databased}. However, for most of practical real systems, the existence of external disturbances is usually unavoidable. 

To reduce the effects of disturbance, robust controller is required for disturbance rejection. One effective solution is the $ H_\infty $ control method, which achieves disturbance attenuation in the $ L_2 $-gain setting \cite{zhou1996robust, schaft1996l2, bacsar2008h}, that is, to design a controller such that the ratio of the objective output energy to the disturbance energy is less than a prescribed level. Over the past few decades, a large number of theoretical results on nonlinear $ H_\infty $ control have been reported \cite{schaft1992l2, isidori1995h, isidori1992disturbance}, where the $ H_\infty $ control problem can be transformed into how to solve the so-called Hamilton-Jacobi-Isaacs (HJI) equation. However, the HJI equation is a nonlinear partial differential equation (PDE), which is difficult or impossible to solve, and may not have global analytic solutions even in simple cases. 

Thus, some works have been reported to solve the HJI equation approximately \cite{schaft1992l2, beard1998successive, abu2006policy, vamvoudakis2012online, feng2009game, liu2013neural, sakamoto2008analytical}. In \cite{schaft1992l2}, it was shown that there exists a sequence of policy iterations on the control input such that the HJI equation is successively approximated with a sequence of Hamilton-Jacobi-Bellman (HJB)-like equations. Then, the methods for solving HJB equation can be extended for the HJI equation. In \cite{saridis1979approximation}, the HJB equation was successively approximated by a sequence of linear PDEs, which were solved with Galerkin approximation in \cite{beard1998successive, beard1997galerkin, beard1998approximate}. In \cite{mehraeen2013adaptive}, the successive approximation method was extended to solve the discrete-time HJI equation.  Similar to \cite{beard1998successive}, a policy iteration scheme was developed in \cite{abu2006policy} for the constrained input system. For implementation purpose of this scheme, a neuro-dynamic programming approach was introduced in \cite{abu2008neurodynamic} and an online adaptive method was given in \cite{modares2014online}. This approach suits for the case that the saddle point exists, thus a situation that the smooth saddle point does not exist was considered in \cite{zhang2011iterative}. In \cite{vamvoudakis2012online}, a synchronous policy iteration method was developed, which is the extension of the work \cite{vamvoudakis2010online}. To improve the efficiency for computing the solution of HJI equation, Luo and Wu \cite{luo2013computationally} proposed a computationally efficient simultaneous policy update algorithm (SPUA). In addition, in \cite{huang1995numerical} the solution of the HJI equation was approximated by the Taylor series expansion, and an efficient algorithm was furnished to generate the coefficients of the Taylor series. It is observed that most of these methods \cite{schaft1992l2, beard1998successive, abu2006policy, vamvoudakis2012online, feng2009game, sakamoto2008analytical, abu2008neurodynamic, luo2013computationally, huang1995numerical} are model-based, where the full system model is required. However, the accurate system model is usually unavailable or costly to obtain for many practical systems. Thus, some RL approaches have been proposed for $ H_\infty $ control design of linear systems \cite{wu2013simultaneous, vrabie2011adaptive} and nonlinear systems \cite{wu2012neural} with unknown internal system model. But these methods are on-policy learning approaches \cite{vamvoudakis2012online, wu2013simultaneous, modares2014online, vrabie2011adaptive, wu2012neural, zhang2013near}, where the cost function should be evaluated by using system data generated with policies being evaluated. It is found that there are several drawbacks (to be discussed in Section \ref{Sec_3}) to apply the on-policy learning to solve real $ H_\infty $ control problem. 

To overcome this problem, this paper introduces an off-policy RL method to solve the nonlinear continuous-time $ H_\infty $ control problem with unknown internal system model. The rest of the paper is rearranged as follows. Sections \ref{Sec_2} and \ref{Sec_3} present the problem description and the motivation. The off-policy learning methods for nonlinear systems and linear systems are developed in \ref{Sec_4} and \ref{Sec_5} respectively. The simulation studies are conducted in Section \ref{Sec_6} and a brief conclusion is given in Section \ref{Sec_7}. 

\textit{Notations}: $\mathbb{R}, \mathbb{R}^n$ and $\mathbb{R}^{n\times m} $  are the set of real numbers, the $ n $-dimensional Euclidean space and the set of all real   matrices, respectively.  $ \Vert \cdot \Vert $ denotes the vector norm or matrix norm in $ \mathbb{R}^n$ or $\mathbb{R}^{n\times m} $   , respectively. The superscript $ T $ is used for the transpose and  $ I $ denotes the identify matrix of appropriate dimension. $ \bigtriangledown \triangleq \partial / \partial x $  denotes a gradient operator notation. For a symmetric matrix $ M, M>(\geqslant) 0$  means that it is a positive (semi-positive) definite matrix.  $ \Vert v \Vert ^2_M \triangleq v^T M v $  for some real vector  $ v $ and symmetric matrix    $ M>(\geqslant) 0$ with appropriate dimensions. $ C^1(\mathcal{X}) $  is function space on $ \mathcal{X} $  with first derivatives are continuous.  $ L_2 [0,\infty) $  is a Banach space, for $ \forall w(t) \in L_2 [0,\infty),  \int_{0}^{\infty} \Vert w(t) \Vert ^2 dt < \infty$. Let $ \mathcal{X},\mathcal{U} $ and $ \mathcal{W} $ be compact sets, denote $ \mathcal {D}  \triangleq \lbrace (x,u,w) | x \in \mathcal{X}, u \in \mathcal{U}, w \in \mathcal{W} \rbrace $. For column vector functions $ s_1(x,u,w) $ and $ s_2(x,u,w) $  , where $ (x,u,w) \in \mathcal {D} $ Ôºå define inner product $ \langle s_1(x,u,w), s_2(x,u,w)\rangle_{\mathcal {D} } \triangleq \int_{\mathcal {D} } s_1^T(x,u,w) s_2(x,u,w) d(x,u,w) $  and norm  $ \Vert s_1(x,u,w) \Vert _{\mathcal {D} } $ $ \triangleq \left( \int_{\mathcal {D} } s_1^T(x,u,w) s_1(x,u,w) d(x,u,w) \right)^{1/2}$. $ H^{m,p}(\mathcal{X}) $ is a Sobolev space that consists of functions in space $ L_p (\mathcal{X}) $ such that their derivatives of order at least $m$  are also in $ L_p (\mathcal{X}) $.

\section{Problem description} \label{Sec_2}
Let us consider the following affine nonlinear continuous-time dynamical system:
\begin{flalign}
\dot{x}(t) &= f(x(t)) + g(x(t))u(t) + k(x(t))w(t) \label{eq_sys}\\
z(t) &= h(x) \label{eq_h(x)}
\end{flalign}
where $[x_1~ ... ~x_n]^T \in \mathcal{X} \subset \mathbb{R}^n$ is the state,   $u = [u_1~ ... ~u_m]^T \in \mathcal{U} \subset \mathbb{R}^m$ is the control input and $ u(t) \in L_2 [0,\infty) $, $ w = [w_1~ ... ~w_q]^T \in \mathcal{W} \subset \mathbb{R}^q$ is the external disturbance and $ w(t) \in L_2 [0,\infty) $, $ z = [z_1~ ... ~z_p]^T \in \mathbb{R}^p$ is the objective output. $ f(x) $  is Lipschitz continuous on a set $\mathcal{X}$  that contains the origin, $ f(0)=0 $. $ f(x) $ represents the internal system model which is assumed to be \textit{unknown} in this paper. $ g(x), k(x) $ and $ h(x) $  are known continuous vector or matrix functions of appropriate dimensions.

The $ H_\infty $ control problem under consideration is to find a state feedback control law  $ u(x) $ such that the system \eqref{eq_sys} is closed-loop asymptotically stable, and has  $ L_2 $-gain less than or equal to $ \gamma $ , that is,
\begin{equation} \label{L2_gain}
\int_{0}^{\infty} {\left( \Vert z(t) \Vert ^2 + \Vert u(t) \Vert _R^2 \right)} dt \leq \gamma^2 \int_{0}^{\infty} {\Vert w(t) \Vert ^2} dt
\end{equation}
for all $ w(t) \in L_2 [0,\infty), R>0 $ and $ \gamma>0 $ is some prescribed level of disturbance attenuation. From \cite{schaft1992l2}, this problem can be transformed to solve the so-called HJI equation, which is summarized in Lemma \ref{lemma_1}. 

\begin {lemma} \label{lemma_1}
Assume the system \eqref{eq_sys} and \eqref{eq_h(x)} is zero-state observable. For $ \gamma>0 $ , suppose there exists a solution $V^*(x) $  to the HJI equation
\begin{flalign} \label{HJI}
&[\nabla V^*(x)]^T f(x) + h^T(x)h(x) \nonumber \\
&\qquad- \frac{1}{4} [\nabla V^*(x)]^T  g(x) R^{-1} g^T(x) \nabla V^*(x) \nonumber \\
&\qquad + \frac{1}{4\gamma^2} [\nabla V^*(x)]^T  k(x) k^T(x) \nabla V^*(x) = 0.
\end{flalign} 
where $ V^*(x) \in C^1(\mathcal{X}), V^*(x) \geqslant 0 $ and $ V^*(0) = 0 $. Then, the closed-loop system with the state feedback control
\begin{equation}\label{idea_optimal_control}
u(t) = u^*(x(t)) = -\frac{1}{2} R^{-1} g^T(x) \nabla V^*(x)
\end{equation}
has $ L_2 $-gain less than or equal to $ \gamma $, and the closed-loop system \eqref{eq_sys}, \eqref{eq_h(x)} and \eqref{idea_optimal_control} (when $ w(t)\equiv 0 $ ) is locally asymptotically stable. $\square$
\end {lemma}

\section{Motivation from investigation of related work} \label{Sec_3}
From Lemma \ref{lemma_1}, it is noted that the $H_\infty$ control \eqref{idea_optimal_control} relies on the solution of the HJI equation \eqref{HJI}. Therefore, a model-based iterative method was proposed in \cite{beard1998successive}, where the HJI equation is successively approximated by a sequence of linear PDEs:  
\begin{eqnarray}\label{eq_3.1}
[\nabla V^{(i,j+1)}]^T (f + gu^{(i)} + kw^{(i,j)}) + h^Th + \Vert u^{(i)} \Vert _R^2 \nonumber \\
- \gamma^2 \Vert w^{(i,j)} \Vert ^2=0
\end{eqnarray}
and then update control and disturbance policies with
\begin{flalign}
w^{(i,j+1)} & \triangleq \frac{1}{2} \gamma^{-2} k^T \nabla V^{(i,j+1)}\label{eq_3.2}  \\
u^{(i+1)} & \triangleq -\frac{1}{2} R^{-1} g^T \nabla V^{(i+1)}\label{eq_3.3}
\end{flalign}
with $ V^{(i+1)} \triangleq \lim\limits_{j \to \infty} V^{(i, j)}$. From \cite{schaft1992l2, beard1998successive}, it was indicated that the $ V^{(i,j)} $ can converge to the solution of the HJI equation, i.e., $\lim\limits_{i,j \to \infty} V^{(i, j)} = V^* $. 

\begin{remark} \label{remark1}
\emph{Note that the key point of the iterative scheme \eqref{eq_3.1}-\eqref{eq_3.3} depends on the solution of the linear PDE \eqref{eq_3.1}. Thus, several related methods were developed, such as, Galerkin approximation  \cite{beard1998successive}, synchronous policy iteration \cite{vamvoudakis2012online}, neuro-dynamic programming approach \cite{abu2006policy, abu2008neurodynamic} and online adaptive control method \cite{modares2014online} for constrained input systems, and Galerkin approximation method for discrete-time systems \cite{mehraeen2013adaptive}. Obviously, the iteration \eqref{eq_3.1}-\eqref{eq_3.3} will generate two iterative loops since the control and disturbance policies are updated at the different iterative steps, i.e., the inner loop for updating disturbance policy $w$ with index $j$, and the outer iterative loop for updating control policy $u$ with index $i$. The outer loop will not be activated until the inner loop is convergent, which results in low efficiency. Therefore, Luo and Wu \cite{luo2013computationally} proposed a simultaneous policy update algorithm (SPUA), where the control and disturbance policies are updated at the same iterative step, and thus only one iterative loop is required. 
It worth noting that the word ``simultaneous" in \cite{luo2013computationally} and the word ``synchronous/simultaneous" in \cite{vamvoudakis2012online,modares2014online} represent different meanings. The former emphasizes the same ``iterative step,‚Äù while the latter emphasizes the same ``time instant". In other words, the SPUA in \cite{luo2013computationally} updates control and disturbance policies at the ``same" iterative step, while the algorithms in \cite{vamvoudakis2012online,modares2014online}  update control and disturbance policies at the ``different" iterative steps.
$ \square $}
\end{remark}

The procedure of model-based SPUA is given in Algorithm \ref{algorithm_1}.

\noindent \rule{0.49\textwidth}{2pt}
\begin{algorithm} \label{algorithm_1}
\emph{
Model-based SPUA.
\begin{itemize}
\item [$\blacktriangleright$] \emph{ Step 1:} Give an initial function $ V^{(0)} \in \mathbb{V}_0 $ ($ \mathbb{V}_0 \subset \mathbb{V} $ is determined by Lemma 5 in \cite{luo2013computationally}. Let $ i=0 $ ; 
\item [$\blacktriangleright$] \emph{Step 2:} Update the control and disturbance policies with
\begin{flalign}
u^{(i)} & \triangleq -\frac{1}{2} R^{-1} g^T \nabla V^{(i)}\label{eq_control_update}\\
w^{(i)} & \triangleq \frac{1}{2} \gamma^{-2} k^T \nabla V^{(i)}\label{eq_distur_update}
\end{flalign} 
\item [$\blacktriangleright$] \emph{Step 3:} Solve the following linear PDE for $ V^{(i+1)}(x) $ :
\begin{eqnarray}\label{LFE}
[\nabla V^{(i+1)}]^T (f + gu^{(i)} + kw^{(i)}) + h^Th + \Vert u^{(i)} \Vert _R^2 \nonumber \\
- \gamma^2 \Vert w^{(i)} \Vert ^2=0;
\end{eqnarray}
where $ V^{(i+1)}(x) \in C^1(\mathcal{X}), V^{(i+1)}(x) \geqslant 0 $ and $ V^{(i+1)}(0) = 0 $. 
\item [$\blacktriangleright$] \emph{Step 4:}   Let $ i = i+1 $, go back to Step 2 and continue. $\square $ 
\end{itemize} 
}
\end{algorithm}
\noindent \rule{0.49\textwidth}{2pt}

It worth noting that Algorithm \ref{algorithm_1} is an infinite iterative procedure, which is used for theoretical analysis rather than for implementation purpose. That is to say, Algorithm \ref{algorithm_1} will converge to the solution of the HJI equation \eqref{HJI} when the iteration goes to infinity. 
By constructing a fixed point equation, the convergence of Algorithm \ref{algorithm_1} is established in \cite{luo2013computationally} by proving that it is essentially a Newton‚Äôs iteration method for finding the fixed point. With the increase of index $i$, the sequence $ V^{(i)} $ obtained by the SPUA with equations \eqref{eq_control_update}-\eqref{LFE} can converge to the solution of HJI equation \eqref{HJI}, i.e., $\lim\limits_{i \to \infty} V^{(i)} = V^* $.
\begin{remark} \label{remark2a}
\emph{It is necessary to explain the rationale of using equations \eqref{eq_control_update} and \eqref{eq_distur_update} for control and disturbance policies update. The $ H_\infty $ control problem \eqref{eq_sys}-\eqref{L2_gain} can be viewed as a two-players zero-sum differential game problem \cite{bacsar2008h,feng2009game, abu2008neurodynamic, vrabie2011adaptive, vamvoudakis2012online, liu2013neural, modares2014online}. The game problem is a minimax problem, where the control policy $ u $ acts as the minimizing player and the disturbance policy $ w $ is the maximizing player. The game problem aims at finding the saddle point $ (u^*,w^* )$, where $ u^* $ is given by expression \eqref{idea_optimal_control} and $ w^* $ is given by $ w^*(x)  \triangleq \frac{1}{2} \gamma^{-2} k^T(x) \nabla V^*(x) $. Correspondingly, for the $ H_\infty $ control problem \eqref{eq_sys}-\eqref{L2_gain}, $ u^* $ and $ w^* $ are the associated $ H_\infty $ control policy and the worst disturbance signal \cite{bacsar2008h, abu2006policy, abu2008neurodynamic, vrabie2011adaptive, vamvoudakis2012online, liu2013neural}, respectively. Thus, it is reasonable using expressions \eqref{eq_control_update} and \eqref{eq_distur_update} (that are consistent with $ u^* $ and $ w^* $ in form) for control and disturbance policies update.
Similar control and disturbance policy update method could be found in references \cite{schaft1992l2, beard1998successive, abu2006policy, abu2008neurodynamic, liu2013neural, modares2014online}. 
$ \square $}
\end{remark}

Observe that both iterative equations \eqref{eq_3.1} and \eqref{LFE} require the full system model. For the $H_\infty$ control problem that the internal system dynamic $ f(x) $ is unknown, data based methods \cite{vrabie2011adaptive, wu2012neural} were suggested to solve the HJI equation online. However, most of related existing online methods are on-policy learning approaches \cite{vamvoudakis2012online, modares2014online, vrabie2011adaptive, wu2012neural, zhang2013near}. From the definition of on-policy learning \cite{sutton1998reinforcement}, the cost function should be evaluated with the data generated from the evaluating policies. For example,  $ V^{(i,j+1)} $ in equation \eqref{eq_3.1} is the cost function of the policies  $ w^{(i,j)} $ and $ u^{(i)} $, which means that $ V^{(i,j+1)} $ should be evaluated with system data by using evaluating policies $ w^{(i,j)} $ and $ u^{(i)} $. It is observed that these on-policy learning approaches for solving the $H_\infty$ control problem have several drawbacks: 
\begin{itemize}
\item 1) For real implementation of on-policy learning methods \cite{vamvoudakis2012online, modares2014online, wu2012neural, zhang2013near}, the approximate evaluating control and disturbance policies (rather than the actual policies) are used to generate data for learning their cost function. In other words, the on-policy learning methods using the ``inaccurate" data to learn their cost function, which will increase the accumulated error. For example, to learn the cost function $ V^{(i,j+1)} $ in equation \eqref{eq_3.1}, some approximate policies  $ \widehat{w}^{(i,j)} $ and $ \widehat{u}^{(i)} $ (rather than its actual policies  $ w^{(i,j)} $ and $ u^{(i)} $, which are usually unknown because of estimate error) are employed to generate data; 
\item 2) The evaluating control and disturbance policies are required to generate data for on-policy learning, thus disturbance signal should be adjustable, which is usually impractical for most of real systems; 
\item 3) It is known \cite{sutton1998reinforcement, thrun1992efficient} that the issue of ``exploration" is extremely important in RL for learning the optimal control policy, and the lack of exploration during the learning process may lead to divergency. Nevertheless, for on-policy learning, exploration is restricted because only the evaluating policies can be used to generate data. From the literature investigation, it is found that the ``exploration" issue is rarely discussed in existing work that using RL techniques for control design; 
\item 4) The implementation structure is complicated, such as in  \cite{vamvoudakis2012online, modares2014online}, three NNs are required for approximating cost function, control and disturbance policies, respectively; 
\item 5) Most of existing approaches \cite{vamvoudakis2012online, modares2014online, vrabie2011adaptive, wu2012neural, zhang2013near} are implemented online, thus they are difficult for real-time control because the learning process is often time-consuming. Furthermore, online control design approaches just use current data while discard past data, which implies that the measured system data is used only once and thus results in low utilization efficiency.
\end{itemize}
To overcome the drawbacks mentioned above, we propose an off-policy RL approach to solve the $H_\infty$ control problem with unknown internal system dynamic $ f(x) $.  

\section{Off-policy reinforcement learning for $ H_\infty $ control} \label{Sec_4}
In this section, an off-policy RL method for $H_\infty$ control design is derived and its convergence is proved. Then, a NN-based critic-actor structure is developed for implementation purpose.
 
\subsection{Off-policy reinforcement learning} \label{}
To derive the off-policy RL method, we rewrite the system \eqref{eq_sys} as:
\begin{equation}\label{eq_4_1}
\dot{x} = f + gu^{(i)} + kw^{(i)} + g[u - u^{(i)}] + k[ w-w^{(i)} ].
\end{equation}
for $ \forall u \in \mathcal{U},  w \in \mathcal{W} $. Let $ V^{(i+1)}(x) $  be the solution of the linear PDE \eqref{LFE}, then taking derivative along the state of system \eqref{eq_4_1} yields, 
\begin{flalign}\label{eq_4_2}
&\frac{dV^{(i+1)}(x)}{dt} = [\nabla V^{(i+1)}]^T (f + gu^{(i)} + kw^{(i)}) \nonumber \\
&\quad + [\nabla V^{(i+1)}]^T g[u - u^{(i)}] + [\nabla V^{(i+1)}]^T k[ w-w^{(i)} ].
\end{flalign}
With the linear PDE \eqref{LFE}, conducting integral on both sides of equation \eqref{eq_4_2} in time interval $ [t, t+\Delta t] $ and rearranging terms yield,
\begin{flalign}
& \int_{t}^{t+\Delta t} [\nabla V^{(i+1)}(x(\tau))]^T g(x(\tau))[u(\tau) - u^{(i)}(x(\tau))] d \tau \nonumber \\
& \quad + \int_{t}^{t+\Delta t} [\nabla V^{(i+1)}(x(\tau))]^T k(x(\tau))[ w(\tau)-w^{(i)}(x(\tau)) ] d \tau \nonumber \\
& \quad +V^{(i+1)}(x(t)) - V^{(i+1)}(x(t+\Delta t)) \nonumber
\end{flalign}
\begin{flalign}\label{eq_4_3}
& = \int_{t}^{t+\Delta t}  \left( h^T(x(\tau))h(x(\tau)) + \Vert u^{(i)} (x(\tau)) \Vert _R^2 \right. \nonumber\\
& \quad \left. - \gamma^2 \Vert w^{(i)} (x(\tau)) \Vert ^2 \right) d\tau
\end{flalign} 
It is observed from the equation \eqref{eq_4_3} that the cost function $ V^{(i+1)} $ can be learned by using arbitrary input signals $ u $ and $ w $, rather than the evaluating  policies $ u^{(i)} $ and $ w^{(i)} $. Then, 
replacing linear PDE \eqref{LFE} in Algorithm \ref{algorithm_1} with \eqref{eq_4_3} results in the off-policy RL method. To show its convergence, Theorem \ref{theorem_4.1} establishes the equivalence between iterative equations \eqref{LFE} and \eqref{eq_4_3}. 

\begin{theorem}\label{theorem_4.1}
Let $ V^{(i+1)}(x) \in C^1(\mathcal{X}), V^{(i+1)}(x) \geqslant 0 $ and $ V^{(i+1)}(0) = 0 $. $ V^{(i+1)}(x) $ is the solution of equation \eqref{eq_4_3} iff $($if and only if $)$ it is the solution of the linear PDE \eqref{LFE}, i.e., equation \eqref{eq_4_3} is equivalent to the linear PDE \eqref{LFE}.
\end{theorem} 

\noindent \textbf{Proof.} From the derivation of equation \eqref{eq_4_3}, it is concluded that if $ V^{(i+1)}(x) $ is the solution of the linear PDE \eqref{LFE}, then $ V^{(i+1)}(x) $ also satisfies equation \eqref{eq_4_3}. To complete the proof, we have to show that $ V^{(i+1)}(x) $ is the unique solution of equation \eqref{eq_4_3}. The proof is by contradiction.

Before starting the contradiction proof, we derive a simple fact. Consider
\begin{flalign} \label{eq_4_4}
& \lim\limits_{\Delta t \to 0} \frac{1}{\Delta t} \int_{t}^{t+\Delta t} \hbar (\tau) d \tau \nonumber \\
& \quad = 
\lim\limits_{\Delta t \to 0} \frac{1}{\Delta t} \left( \int_{0}^{t+\Delta t} \hbar (\tau) d \tau - \int_{0}^{t} \hbar (\tau) d \tau \right) \nonumber \\
& \quad =  \frac{d}{dt} \int_{0}^{t} \hbar (\tau) d \tau \nonumber \\
& \quad =   \hbar (t).
\end{flalign}

From \eqref{eq_4_3}, we have
\begin{flalign}\label{eq_4_5}
& \frac{dV^{(i+1)}(x)}{dt} =  \lim\limits_{\Delta t \to 0} \frac{1}{\Delta t} \left[ V^{(i+1)}(x(t + \Delta t)) - V^{(i+1)}(x(t)) \right] \nonumber\ \\ 
&\quad  = \lim\limits_{\Delta t \to 0} \frac{1}{\Delta t}  \int_{t}^{t+\Delta t} [\nabla V^{(i+1)}(x(\tau))]^T g(x(\tau)) \nonumber \\
&\qquad  [u(\tau) - u^{(i)}(x(\tau))] d \tau \nonumber \\
&\quad + \lim\limits_{\Delta t \to 0} \frac{1}{\Delta t} \int_{t}^{t+\Delta t} [\nabla V^{(i+1)}(x(\tau))]^T k(x(\tau))\nonumber \\ 
&\qquad [ w(\tau)-w^{(i)}(x(\tau)) ] d \tau \nonumber \\ 
&\quad - \lim\limits_{\Delta t \to 0} \frac{1}{\Delta t} \int_{t}^{t+\Delta t}  \left[ h^T(x(\tau))h(x(\tau)) + \Vert u^{(i)} (x(\tau)) \Vert _R^2 \right. \nonumber \\ 
&\qquad \left. - \gamma^2 \Vert w^{(i)} (x(\tau)) \Vert ^2 \right] d\tau. 
\end{flalign}
By using the fact \eqref{eq_4_4}, the equation \eqref{eq_4_5} is rewritten as
\begin{flalign}\label{eq_4_6}
&\frac{dV^{(i+1)}(x)}{dt} = [\nabla V^{(i+1)}(x(t))]^T g(x(t))[u(t) - u^{(i)}(x(t))]\nonumber \\
&\quad + [\nabla V^{(i+1)}(x(t))]^T k(x(t))[ w(t)-w^{(i)}(x(t)) ]\nonumber \\ 
&\quad - \left[ h^T(x(t))h(x(t)) + \Vert u^{(i)} (x(t)) \Vert _R^2 - \gamma^2 \Vert w^{(i)} (x(t)) \Vert ^2 \right]. 
\end{flalign}
Suppose that $ W(x) \in C^1(\mathcal{X}) $ is another solution of equation \eqref{eq_4_3} with boundary condition $ W(0) = 0 $. Thus, $ W(x) $ also satisfies equation \eqref{eq_4_6}, i.e.,
\begin{flalign}\label{eq_4_7}
& \frac{dW(x)}{dt} = [\nabla W(x(t))]^T g(x(t))[u(t) - u^{(i)}(x(t))] \nonumber \\
&\quad + [\nabla W(x(t))]^T k(x(t))[ w(t)-w^{(i)}(x(t)) ]\nonumber \\ 
&\quad - \left[ h^T(x(t))h(x(t)) + \Vert u^{(i)} (x(t)) \Vert _R^2 - \gamma^2 \Vert w^{(i)} (x(t)) \Vert ^2 \right]. 
\end{flalign}
Substituting equation \eqref{eq_4_7} from \eqref{eq_4_6} yields,
\begin{flalign} \label{eq_4_7a}
&\frac{d}{dt} \left(V^{(i+1)}(x) -W(x) \right) \nonumber \\
&\quad = \left[ \nabla \left(V^{(i+1)}(x) -W(x) \right) \right]^T g(x)[u - u^{(i)}(x)] \nonumber \\
&\quad +\left[ \nabla \left(V^{(i+1)}(x) -W(x) \right) \right]^T k(x)[ w -w^{(i)}(x) ]. 
\end{flalign}
This means that equation \eqref{eq_4_7a} holds for $ \forall u \in \mathcal{U},  w \in \mathcal{W} $. If letting $ u = u^{(i)}, w =w^{(i)} $,  we have
\begin{equation} \label{eq_4_7b}
\frac{d}{dt} \left[ V^{(i+1)}(x) -W(x) \right] = 0. 
\end{equation}
Then, $ V^{(i+1)}(x) -W(x) = c $ for $ \forall x \in \mathcal{X} $, where $ c $  is a real constant, and $ c = V^{(i+1)}(0) -W(0) = 0 $ . Thus,$ V^{(i+1)}(x) -W(x) = 0 $, i.e., $  W(x) = V^{(i+1)}(x) $   for $ \forall x \in \mathcal{X} $. This completes the proof.
$ \square $

\begin{remark} \label{remark3}
\emph{ It follows from Theorem \ref{theorem_4.1} that the solution of equation \eqref{eq_4_3} is equivalent to equation \eqref{LFE}, and thus the convergence of the off-policy RL is guaranteed, i.e., the solution of the iterative equation \eqref{eq_4_3} will converge to the solution of HJI equation \eqref{HJI} as iteration step $i$ increases. Different from the equation \eqref{LFE} in Algorithm \ref{algorithm_1}, the off-policy RL with equation \eqref{eq_4_3} uses system data instead of the internal system dynamic $ f(x) $. Hence, the off-policy RL can be regarded as a direct learning method for $H_\infty$ control design, which avoids the identification of $ f(x) $. In fact, the information of $ f(x) $ is embedded in the measurement of system data. That is to say, the lack of knowledge about $ f(x) $ does not have any impact on the off-policy RL to obtain the solution of HJI equation \eqref{HJI} and the  $H_\infty$ control policy.
It worth pointing out that the equation \eqref{eq_4_3} is similar with the form of the IRL \cite{vrabie2009adaptive, vrabie2009neural}, which is an important framework for control design of continuous-time systems. The IRL in \cite{vrabie2009adaptive, vrabie2009neural} is an online optimal control learning algorithm for partially unknown systems.
$ \square $ }
\end{remark}

\subsection{Implementation based on neural network} \label{}
To solve equation \eqref{eq_4_3} for the unknown function $ V^{(i+1)}(x) $ based on system data, we develop a NN based actor-critic structure. From the well known high-order Weierstrass approximation theorem \cite{courant2004methods}, a continuous function can be represented by an infinite-dimensional linearly independent basis function set. For real practical application, it is usually required to approximate the function in a compact set with a finite-dimensional function set.  We consider the critic NN for approximating the cost function on a compact set $\mathcal{X}$. Let   $\varphi(x) \triangleq [\varphi_1(x)~...~\varphi_L(x)]^T$ be the vector of linearly independent activation functions for critic NN, where  $\varphi_l(x): \mathcal{X}\mapsto \mathbb{R}, l=1,...,L, L$ is the number of critic NN hidden layer neurons. Then, the output of critic NN is given by
\begin{equation}\label{eq_4_8}
\widehat{V}^{(i)} (x) = \sum_{l=1}^{L} \theta_{l}^{(i)} \varphi_l(x) = \varphi^T(x) \theta^{(i)} 
\end{equation}
for $ \forall i=0,1,2,... $, where $\theta^{(i)} \triangleq [\theta_{1}^{(i)}~...~\theta_{L}^{(i)}]^T$ is the critic NN weight vector. It follows from \eqref{eq_control_update}, \eqref{eq_distur_update} and \eqref{eq_4_8} that the disturbance and control policies are given by:
\begin{flalign}
\widehat{u}^{(i)} (x) &= -\frac{1}{2} R^{-1} g^T (x) \nabla \varphi^T(x) \theta^{(i)}\label{eq_4_10}
\\
\widehat{w}^{(i)} (x) &= \frac{1}{2} \gamma^{-2} k^T (x) \nabla \varphi^T(x) \theta^{(i)} \label{eq_4_9} 
\end{flalign}
for $ \forall i=0,1,2,... $,  and $ \nabla \varphi(x) \triangleq [\partial \varphi_1/\partial x~...~\partial \varphi_L/\partial x]^T$ is the Jacobian of $ \varphi(x) $.
Expressions \eqref{eq_4_10} and \eqref{eq_4_9} can be viewed as actor NNs for the disturbance and control policies respectively, where $ -\frac{1}{2} R^{-1} g^T (x) \nabla \varphi^T(x) $ and $ \frac{1}{2} \gamma^{-2} k^T (x) \nabla \varphi^T(x) $ are the activation function vectors and $ \theta^{(i)} $ is the actor NN weight vector.

Due to estimation errors of the critic and actor NNs \eqref{eq_4_8}-\eqref{eq_4_9}, the replacement of $ V^{(i+1)}, w^{(i)} $ and  $ u^{(i)}  $ in the iterative equation \eqref{eq_4_3} with $ \widehat{V}^{(i+1)}, \widehat{w}^{(i)} $ and  $ \widehat{u}^{(i)}  $ respectively, yields the following residual error:
\begin{flalign} \label{eq_4_11}
& \sigma ^{(i)} (x(t), u(t), w(t)) \nonumber\\
& \triangleq \int_{t}^{t+\Delta t} [u(\tau) - \widehat{u}^{(i)}(x(\tau))]^T g^T(x(\tau)) \nabla \varphi^T(x(\tau)) \theta^{(i+1)} d \tau\nonumber\\
& + \int_{t}^{t+\Delta t} [ w(\tau)  -\widehat{w}^{(i)}(x(\tau)) ]^T k^T(x(\tau)) \nabla \varphi^T(x(\tau)) \theta^{(i+1)} d \tau \nonumber\\
& + [\varphi(x(t)) - \varphi(x(t+\Delta t))]^T \theta^{(i+1)} \nonumber\\
& - \int_{t}^{t+\Delta t} \left[ h^T(x(\tau))h(x(\tau)) + \Vert \widehat{u}^{(i)} (x(\tau)) \Vert _R^2 \right.
\nonumber\\
& \quad \left. - \gamma^2 \Vert \widehat{w}^{(i)} (x(\tau)) \Vert ^2 \right] d\tau \nonumber \\
& = \int_{t}^{t+\Delta t} u^T(\tau) g^T(x(\tau)) \nabla \varphi^T(x(\tau)) \theta^{(i+1)} d \tau \nonumber \\
& + \frac{1}{2} \int_{t}^{t+\Delta t} \left( \theta^{(i)} \right)^T \nabla \varphi(x(\tau))  g(x(\tau)) R^{-1} g^T(x(\tau)) \nonumber\\
& \quad \nabla \varphi^T(x(\tau)) \theta^{(i+1)} d \tau \nonumber\\
& + \int_{t}^{t+\Delta t} w^T(\tau) k^T(x(\tau)) \nabla \varphi^T(x(\tau)) \theta^{(i+1)} d \tau \nonumber\\ 
& - \frac{1}{2} \gamma^{-2} \int_{t}^{t+\Delta t} \left( \theta^{(i)} \right)^T \nabla \varphi(x(\tau))  k(x(\tau)) k^T(x(\tau)) \nonumber \\
& \quad \nabla \varphi^T(x(\tau)) \theta^{(i+1)} d \tau + [\varphi(x(t)) - \varphi(x(t+\Delta t))]^T \theta^{(i+1)} \nonumber\\
& - \frac{1}{4} \int_{t}^{t+\Delta t} \left( \theta^{(i)} \right)^T \nabla \varphi(x(\tau))  g(x(\tau)) R^{-1} g^T(x(\tau)) \nonumber\\
&\quad \nabla \varphi^T(x(\tau)) \theta^{(i)} d \tau \nonumber\\
& + \frac{1}{4} \gamma^{-2} \int_{t}^{t+\Delta t} \left( \theta^{(i)} \right)^T \quad \nabla \varphi(x(\tau))  k(x(\tau)) k^T(x(\tau)) \nonumber\\ 
& \quad \nabla \varphi^T(x(\tau)) \theta^{(i)} d \tau - \int_{t}^{t+\Delta t}  h^T(x(\tau))h(x(\tau)) d\tau 
\end{flalign}
For notation simplicity, define
\begin{flalign}
\rho _ {\Delta \varphi} (x(t)) \triangleq &  \left[ \varphi(x(t)) - \varphi(x(t+\Delta t)) \right]^T \nonumber\\
\rho _ {g \varphi} (x(t)) \triangleq & \int_{t}^{t+\Delta t} \nabla \varphi(x(\tau))  g(x(\tau)) R^{-1} g^T(x(\tau)) \nonumber\\
& \nabla \varphi^T(x(\tau)) d\tau \nonumber\\
\rho _ {k \varphi} (x(t)) \triangleq & \int_{t}^{t+\Delta t} \nabla \varphi(x(\tau))  k(x(\tau)) k^T(x(\tau)) \nonumber\\
& \nabla \varphi^T(x(\tau)) d\tau \nonumber\\ 
\rho _ {u \varphi}(x(t),u(t)) \triangleq & \int_{t}^{t+\Delta t} u^T(\tau) g^T(x(\tau)) \nabla \varphi^T(x(\tau)) d \tau \nonumber \\
\rho _ {w \varphi}(x(t),w(t)) \triangleq & \int_{t}^{t+\Delta t} w^T(\tau) k^T(x(\tau)) \nabla \varphi^T(x(\tau)) d \tau \nonumber\\
\rho _ h (x(t)) \triangleq & \int_{t}^{t+\Delta t}  h^T(x(\tau))h(x(\tau)) d\tau \nonumber
\end{flalign}
then, expression \eqref{eq_4_11} is rewritten as
\begin{flalign}\label{eq_4_12}
& \sigma ^{(i)} (x(t), u(t), w(t)) \nonumber\\
& =  \rho _ {u \varphi}(x(t),u(t)) \theta^{(i+1)} 
+ \frac{1}{2} \left( \theta^{(i)} \right)^T \rho _ {g \varphi} (x(t))\theta^{(i+1)} \nonumber\\
& +  \rho _ {w \varphi}(x(t),w(t)) \theta^{(i+1)}  
- \frac{1}{2} \gamma^{-2} \left( \theta^{(i)} \right)^T \rho _ {k \varphi} (x(t)) \theta^{(i+1)} \nonumber\\
& + \rho _ {\Delta \varphi}  \theta^{(i+1)} 
- \frac{1}{4} \left( \theta^{(i)} \right)^T \rho _ {g \varphi} (x(t))\theta^{(i)}  \nonumber\\
& + \frac{1}{4} \gamma^{-2} \left( \theta^{(i)} \right)^T \rho _ {k \varphi} (x(t)) \theta^{(i)} 
-\rho _ h (x(t)).
\end{flalign}
For description convenience, expression \eqref{eq_4_12} is represented as a compact form
\begin{equation} \label{eq_4_13}
\sigma ^{(i)} (x(t), u(t), w(t)) = \overline{\rho}^{(i)}(x(t), u(t), w(t)) \theta^{(i+1)} -\pi^{(i)}(x(t))
\end{equation}
where
\begin{flalign}
& \overline{\rho}^{(i)}(x(t), u(t), w(t)) \triangleq \rho _ {u \varphi}(x(t),u(t)) + \frac{1}{2} \left( \theta^{(i)} \right)^T \rho _ {g \varphi} (x(t)) \nonumber\\
&\quad +  \rho _ {w \varphi}(x(t),w(t)) - \frac{1}{2} \gamma^{-2} \left( \theta^{(i)} \right)^T \rho _ {k \varphi} (x(t))  + \rho _ {\Delta \varphi}  \nonumber\\
& \pi^{(i)}(x(t)) \triangleq \frac{1}{4} \left( \theta^{(i)} \right)^T \rho _ {g \varphi} (x(t))\theta^{(i)} \nonumber\\ 
&\quad - \frac{1}{4} \gamma^{-2} \left( \theta^{(i)} \right)^T \rho _ {k \varphi} (x(t)) \theta^{(i)} 
+ \rho _ h (x(t)) \nonumber.
\end{flalign}

For description simplicity, denote $ \overline{\rho}^{(i)} = [\overline{\rho}_1^{(i)}~...~\overline{\rho}_L^{(i)}]^T$. Based on the method of weighted residuals \cite{finlayson1972method}, the unknown critic NN weight vector  $ \theta^{(i+1)} $ can be computed in such a way that the residual error $ \sigma ^{(i)} (x, u, w) $  (for $ \forall t \geqslant 0 $) of \eqref{eq_4_13} is forced to be zero in some average sense. Thus, projecting the residual error $ \sigma ^{(i)} (x, u, w) $  onto $ d \sigma ^{(i)}/ d \theta^{(i+1)} $  and setting the result to zero on domain $ \mathcal {D} $  using the inner product, $ \langle \cdot , \cdot \rangle _{\mathcal {D} } $, i.e., 
\begin{equation} \label{eq_4_14}
\left<d \sigma ^{(i)}/ d \theta^{(i+1)}, \sigma ^{(i)} (x, u, w) \right> _{\mathcal {D} } = 0.
\end{equation}
Then, the substitution of \eqref{eq_4_13} into \eqref{eq_4_14} yields, 
\begin{eqnarray} 
\left< \overline{\rho}^{(i)}(x, u, w) , \overline{\rho}^{(i)}(x, u, w) \right> _{\mathcal {D} } \theta^{(i+1)} \nonumber \\
-  \left<\overline{\rho}^{(i)}(x, u, w), \pi^{(i)}(x) \right> _{\mathcal {D} } = 0 \nonumber 
\end{eqnarray}
where the notations $ \left< \overline{\rho}^{(i)}, \overline{\rho}^{(i)} \right> _{\mathcal {D} } $ and $ \left<\overline{\rho}^{(i)},  \pi^{(i)} \right> _{\mathcal {D} } $ are given by\\ 
$\left< \overline{\rho}^{(i)}, \overline{\rho}^{(i)} \right> _{\mathcal {D} } \triangleq
\left[ \begin{array}{ccc}
 		\left< \overline{\rho}_1^{(i)}, \overline{\rho}_1^{(i)} \right> _{\mathcal {D} } & \cdots & \left< \overline{\rho}_1^{(i)}, \overline{\rho}_L^{(i)} \right> _{\mathcal {D} } \\
 		\vdots & \cdots & \vdots \\
 		\left< \overline{\rho}_L^{(i)}, \overline{\rho}_1^{(i)} \right> _{\mathcal {D} } & \cdots & \left< \overline{\rho}_L^{(i)}, \overline{\rho}_L^{(i)} \right> _{\mathcal {D} }
	\end{array} \right]$ 
and 
$ \left<\overline{\rho}^{(i)},  \pi^{(i)} \right> _{\mathcal {D} } \triangleq 
\left[ \begin{array}{ccc}
 		\left< \overline{\rho}_1^{(i)}, \pi^{(i)} \right> _{\mathcal {D} }~
 		\cdots~
 		\left< \overline{\rho}_L^{(i)}, \pi^{(i)} \right> _{\mathcal {D} }
	\end{array} \right]^T
$.
\\
Thus, $ \theta^{(i+1)}  $ can be obtained with
\begin{flalign} \label{eq_4_15}
\theta^{(i+1)} = &\left< \overline{\rho}^{(i)}(x, u, w) , \overline{\rho}^{(i)}(x, u, w) \right> ^{-1} _{\mathcal {D} } \nonumber \\
&\left<\overline{\rho}^{(i)}(x, u, w),  \pi^{(i)}(x) \right> _{\mathcal {D} }.
\end{flalign}

The computation of inner products $ \left< \overline{\rho}^{(i)}(x, u, w) , \overline{\rho}^{(i)}(x, u, w) \right> _{\mathcal {D} } $ and $ \left<\overline{\rho}^{(i)}(x, u, w),  \pi^{(i)}(x) \right> _\mathcal {D} $ involve many numerical integrals on domain $ \mathcal {D} $, which are computationally expensive. Thus, the Monte-Carlo integration method \cite{peter1978new} is introduced, which is especially competitive on multi-dimensional domain. We now illustrate the Monte-Carlo integration for computing $ \left< \overline{\rho}^{(i)}(x, u, w) , \overline{\rho}^{(i)}(x, u, w) \right> _{\mathcal {D} } $. Let  $ I_\mathcal {D} \triangleq \int_\mathcal {D} d(x, u, w)  $, and $ \mathcal {S}_M \triangleq \lbrace (x_m, u_m, w_m) \vert (x_m, u_m, w_m) \in \mathcal {D}, m = 1,2,...,M \rbrace  $   be the set that sampled on domain  $ \mathcal {D} $, where $ M $ is size of sample set $ \mathcal {S}_M $. Then, $ \left< \overline{\rho}^{(i)}(x, u, w) , \overline{\rho}^{(i)}(x, u, w) \right> _{\mathcal {D} } $  is approximately computed with
\begin{flalign} \label{eq_4_16}
& \left< \overline{\rho}^{(i)}(x, u, w) , \overline{\rho}^{(i)}(x, u, w) \right> _\mathcal {D} \nonumber \\
&\quad = \int _\mathcal {D} \left( \overline{\rho}^{(i)}(x, u, w) \right)^T \overline{\rho}^{(i)}(x, u, w) d(x,u,w)\nonumber \\
&\quad = \frac{I_\mathcal {D}}{M} \sum _{m=1}^M \left( \overline{\rho}^{(i)}(x_m, u_m, w_m) \right)^T \overline{\rho}^{(i)}(x_m, u_m, w_m) \nonumber \\
&\quad = \frac{I_\mathcal {D}}{M} \left( Z^{(i)} \right)^T Z^{(i)}
\end{flalign} 
where $ Z^{(i)} \triangleq \left[ \left( \overline{\rho}^{(i)}(x_1, u_1, w_1) \right)^T~...~\left( \overline{\rho}^{(i)}(x_M, u_M, w_M) \right)^T \right]^T $. Similarly, 
\begin{flalign} \label{eq_4_17}
& \left<\overline{\rho}^{(i)}(x, u, w),  \pi^{(i)}(x) \right> _\mathcal {D} \nonumber \\
&\quad  = \frac{I_\mathcal {D}}{M} \sum _{m=1}^M \left( \overline{\rho}^{(i)}(x_m, u_m, w_m) \right)^T \pi^{(i)}(x_m) \nonumber \\
&\quad = \frac{I_\mathcal {D}}{M} \left( Z^{(i)} \right)^T \eta^{(i)}
\end{flalign}
where $ \eta^{(i)} \triangleq \left[ \pi^{(i)}(x_1)~...~\pi^{(i)}(x_M) \right]^T $. Then, the substitution of \eqref{eq_4_16} and \eqref{eq_4_17} into \eqref{eq_4_15} yields,
\begin{equation} \label{eq_4_18}
\theta^{(i+1)} = \left[ \left( Z^{(i)} \right)^T Z^{(i)} \right]^{-1} \left( Z^{(i)} \right)^T \eta^{(i)}.
\end{equation}

It is noted that the critic NN weight update rule \eqref{eq_4_18} is a least-square scheme. Based on the update rule \eqref{eq_4_18}, the procedure for $H_\infty$ control design with NN-based off-policy RL is presented in Algorithm \ref{algorithm_2}.
\begin{remark} \label{remark4a}
\emph{In the least-square scheme \eqref{eq_4_18}, it is required to compute the inverse of matrix $\left( Z^{(i)} \right)^T Z^{(i)} $. This means that the matrix $ Z^{(i)} $ should be full column rank, which depends on the richness of the sampling data set $S_M$ and its size $M$. To attain this goal in real implementation, it would be useful by increasing the size $M$, starting from different initial states, and using rich input signals, such as random noises, sinusoidal function noises with enough frequencies.
Of course, it would be nice, if possible but is not a necessity, to use the persistent exciting input signals, while it is still a difficult issue \cite{farrell2006adaptive, slotine1991applied} that requires further investigation. In a word, the choices of rich input signals and the size $M$ are generally experience-based. $ \square $ }
\end{remark}

\noindent \rule{0.49\textwidth}{2pt}
\begin{algorithm} \label{algorithm_2}
\emph{
NN-based off-policy RL for $H_\infty$ control design.
\begin{itemize}
\item [$\blacktriangleright$] \emph{Step 1:} Collect real system data $ (x_m, u_m, w_m ) $ for sample set $ \mathcal {S}_M $, and then compute $ \rho _ {\Delta \varphi} (x_m), \rho _ {g \varphi} (x_m), \rho _ {k \varphi} (x_m), $ $\rho _ {u \varphi}(x_m,u_m), $ $\rho _ {w \varphi}(x_m,w_m)  $ and $ \rho _ h (x_m) $;
\item [$\blacktriangleright$] \emph{Step 2:} Select initial critic NN weight vector $ \theta^{(0)} $ such that $ \widehat{V}^{(0)} \in \mathbb{V}_0 $. Let $ i = 0 $;
\item [$\blacktriangleright$] \emph{Step 3:} Compute $ Z^{(i)} $ and $\eta^{(i)} $, and update $\theta^{(i+1)} $ with \eqref{eq_4_18};
\item [$\blacktriangleright$] \emph{Step 4:} Let $i = i+1 $.  If $ \Vert \theta^{(i)} - \theta^{(i-1)} \Vert \leq \xi $ ($ \xi $ is a small positive number), stop iteration and $\theta^{(i)} $ is employed to obtain the $ H_\infty $ control policy with \eqref{eq_4_10}, else go back to Step 3 and continue. $\square $
\end{itemize}
}
\end{algorithm}
\noindent \rule{0.49\textwidth}{2pt}

Note that Algorithm \ref{algorithm_2} has two parts: the first part is Step 1 for data processing, i.e., measure system data $ (x,u,w) $ for computing $ \rho _ {\Delta \varphi}, \rho _ {g \varphi}, \rho _ {k \varphi}, \rho _ {u \varphi}, \rho _ {w \varphi}  $ and $ \rho _ h $; the second part is Steps 2-4 for offline iterative learning the solution of the HJI equation \eqref{HJI}. 
\begin{remark} \label{remark4}
\emph{From Theorem \ref{theorem_4.1}, the proposed off-policy RL is mathematically equivalent to the model-based SPUA (i.e., Algorithm \ref{algorithm_2}), which is proved to be a Newton's method \cite{luo2013computationally}. Hence, the off-policy RL have the same advantages and disadvantages as the Newton's method. That is to say, the off-policy RL is a local optimization method, and thus there exists a problem that an initial critic NN weight vecotr $ \theta^{(0)} $ should be given such that the initial solution $ \widehat{V}^{(0)} $ locates in a neighbourhood $ \mathbb{V}_0 $ of the HJI equation \eqref{HJI}. In fact, this problem also widely arises in many existing works for solving optimal and $H_\infty$ control problems of either linear or nonlinear systems through the observations from computer simulation, such as \cite{kleinman1968iterative, saridis1979approximation, schaft1992l2, beard1998successive, murray2002adaptive, abu2005nearly, abu2006policy, abu2008neurodynamic, chen2008generalized, vrabie2009adaptive, vrabie2009neural, vamvoudakis2011multi,zhang2011iterative}. Till present, it is still a difficult issue for finding proper initializations or developing global approaches. There is no exception for the proposed off-policy RL algorithm, where the selection of initial weight vector $\theta^{(i)} $ is still experience-based and requires further investigation. $ \square $ }
\end{remark}

Algorithm \ref{algorithm_2} can be viewed as an off-policy learning method according to references \cite{sutton1998reinforcement, precup2001off, maei2010toward}, which overcomes the drawbacks mentioned in Section \ref{Sec_3}, i.e.,
\begin{itemize}
\item 1) In the off-policy RL algorithm (i.e., Algorithm \ref{algorithm_2}), 
the control $ u $ and disturbance $ w $ can be arbitrarily on $\mathcal{U} $ and $ \mathcal{W} $, 
where no error occurs during the process of generating data, and thus the accumulated error (exists in the on-policy learning methods mentioned in Section \ref{Sec_3}) can be reduced; 
\item 2) In the Algorithm \ref{algorithm_2}, the control $ u $ and disturbance $ w $ can be arbitrarily on $\mathcal{U} $ and $ \mathcal{W} $, and thus disturbance $ w $ does not required to be adjustable; 
\item 3) In the Algorithm \ref{algorithm_2},  the cost function $ V^{(i+1)} $ of control and disturbance policies $ (u^{(i)} , w^{(i)}) $ can be evaluated by using system data generated with other different control and disturbance signals $ (u, w) $. Thus, the obvious advantage of the developed off-policy RL method is that it can learn the cost function and control policy from system data that are generated according to a more exploratory or even random policies; 
\item 4) The implementation of Algorithm \ref{algorithm_2} is very simple, in fact \textit{only one} NN is required, i.e., critic NN. This means that once the critic NN weight vector $ \theta^{(i+1)}  $ is computed via \eqref{eq_4_18}, the action NNs for control and disturbance policies can be obtained based on \eqref{eq_4_10} and \eqref{eq_4_9} accordingly; 
\item 5) The developed off-policy RL method learns the  $ H_\infty $ control policy offline, which is then used for real-time control. Thus, it is much more practical than online control design methods since less computational load will generate during real-time application. Meanwhile, note that in Algorithm \ref{algorithm_2}, once the terms $ \rho _ {\Delta \varphi}, \rho _ {g \varphi}, \rho _ {k \varphi}, \rho _ {u \varphi}, \rho _ {w \varphi}  $ and $ \rho _ h $ are computed with sample set $ \mathcal {S}_M $ (i.e., Step 1 is finished), no extra data is required for learning the $ H_\infty $ control policy (in Steps 2-4). This means that the collected data set can be utilized repeatedly, and thus the utilization efficiency is improved compared to the online control design methods.
\end{itemize}
\begin{remark} \label{remark5a}
\emph{Observe that the experience replay based IRL method \cite{modares2014integral} can be viewed as an off-policy method based on its definition \cite{sutton1998reinforcement}. There are three obvious differences between the method and the work of this paper. Firstly, the method in \cite{modares2014integral} is for solving the optimal control problem without external disturbance, while the off-policy RL algorithm in this paper is for solving the $ H_\infty $ control problem with external disturbance. Secondly, the method in \cite{modares2014integral} is online adaptive control approach. The off-policy RL algorithm in this paper uses real system information, and  learns the $ H_\infty $ control policy by using an offline  process. After the learning process is finished, the convergent control policy is employed for real system control. Thirdly, the method in \cite{modares2014integral}  involves two NNs (i.e., one critic NN and one actor NN) for adaptive optimal control realization, while only one NN (i.e., critic NN) is required in the algorithm of this paper.$ \square $ }
\end{remark}
\subsection{Convergence analysis for NN-based off-policy RL} \label{}
It is necessary to analyze the convergence of the NN-based off-policy RL algorithm. From Theorem \ref{theorem_4.1}, the equation \eqref{eq_4_3} in the off-policy RL is equivalent to the linear PDE \eqref{LFE}, which means that the derived least-square scheme \eqref{eq_4_18} is essentially for solving the linear PDE \eqref{LFE}. In \cite{abu2005nearly}, a similar least-square method was suggested to solve the first order linear PDE directly, wherein some theoretical results are useful for analyzing the convergence of the proposed NN-based off-policy RL algorithm. The following Theorem \ref{theorem_4.2} is given to show the convergence of critic NN and actor NNs.
\begin{theorem}\label{theorem_4.2}
For $ i=0,1,2,... $, assume that $ V^{(i+1)} \in H^{1,2}(\mathcal{X}) $ is the solution of \eqref{eq_4_3}, the critic NN activation functions $ \varphi_l(x) \in H^{1,2}(\mathcal{X}) $, $ l=1,2,...L $ are selected such that they are complete when $ L \rightarrow \infty $, $ V^{(i+1)} $ and $ \nabla V^{(i+1)} $ can be uniformly approximated, and the set $ \{ \varpi_l (x_1,x_2) \triangleq \varphi_l(x_1) - \varphi_l(x_2) \}_{l=1}^L $ is linearly independent and complete for $ \forall x_1,x_2 \in \mathcal{X}, x_1 \neq x_2$. Then, 
\begin{flalign}
& \sup_{x \in \mathcal{X}} |\widehat{V}^{(i+1)}(x) - V^{(i+1)}(x)| \rightarrow 0 \label{eq_4_3.1} \\
& \sup_{x \in \mathcal{X}} |\nabla \widehat{V}^{(i+1)}(x) - \nabla V^{(i+1)}(x)| \rightarrow 0 \label{eq_4_3.2} \\
& \sup_{x \in \mathcal{X}} |\widehat{u}^{(i+1)}(x) - u^{(i+1)}(x)| \rightarrow 0 \label{eq_4_3.3} \\
& \sup_{x \in \mathcal{X}} |\widehat{w}^{(i+1)}(x) - w^{(i+1)}(x)| \rightarrow 0 \label{eq_4_3.4}.
\end{flalign} 
\end{theorem} 
\noindent \textbf{Proof.} The proof procedure of the above results is very similar with that in reference \cite{abu2005nearly}, and thus some similar proof steps will be omitted for avoidance of repetition. To use the theoretical results in \cite{abu2005nearly}, we firstly prove the $ \{ \nabla \varphi_l (f+gu^{(i)}+kw^{(i)}) \}_{l=1}^L $ is linear independent by contradiction. Assume this is not true, then there exists a vector $\theta \triangleq [\theta_{1}~...~\theta_{L}]^T \neq 0 $ such that
\begin{equation}
\sum_{l=1}^{L} \theta_{l} \nabla \varphi_l (f+gu^{(i)}+kw^{(i)}) = 0 \nonumber \\
\end{equation}
which means that for $ \forall x \in \mathcal{X} $, 
\begin{flalign} 
& \int_{t}^{t+\Delta t} \sum_{l=1}^{L} \theta_{l} \nabla \varphi_l (f+gu^{(i)}+kw^{(i)}) d \tau \nonumber \\
&\quad = \int_{t}^{t+\Delta t} \theta_{l} \frac{d \varphi_l}{d \tau} d \tau \nonumber\\
&\quad = \sum_{l=1}^{L} \theta_{l} [\varphi_l(x(t+\Delta t)) - \varphi_l(x(t))] \nonumber \\
&\quad = \sum_{l=1}^{L} \theta_{l} \varpi_l (x(t+\Delta t),x(t)) \nonumber \\
&\quad = 0. \nonumber 
\end{flalign}
This contradicts the fact that the set $ \{ \varpi_l \}_{l=1}^L $ is linearly independent, which implies that the set $ \{ \nabla \varphi_l (f+gu^{(i)}+kw^{(i)}) \}_{l=1}^L $ is linear independent.\\
\indent From Theorem \ref{theorem_4.1}, $ V^{(i+1)} $ is the solution of  the linear PDE \eqref{LFE}. Then, with the same procedure used in Theorem 2 and Corollary 2 of the reference \cite{abu2005nearly}, the results \eqref{eq_4_3.1}-\eqref{eq_4_3.3} can be proven. And the result \eqref{eq_4_3.4} can be proven in a similar way for \eqref{eq_4_3.3}. $ \square $\\
\indent The results \eqref{eq_4_3.2}-\eqref{eq_4_3.4} in Theorem \ref{theorem_4.2} imply that the critic NN and actor NNs are convergent. In the following Theorem \ref{theorem_4.3}, we prove that the NN-based off-policy RL algorithm converges uniformly to the solution of the HJI equation \eqref{HJI} and the $H_\infty$ control policy \eqref{idea_optimal_control}.
\begin{theorem}\label{theorem_4.3}
If the conditions in Theorem \ref{theorem_4.2} hold, then, for $\forall \epsilon>0$, $ \exists i_0, L_0 $, when $ i \geqslant i_0 $ and $ L \geqslant L_0 $, we have
\begin{flalign}
& \sup_{x \in \mathcal{X}} |\widehat{V}^{(i)}(x) - V^*(x)| < \epsilon \label{eq_4_3.5} \\
& \sup_{x \in \mathcal{X}} |\widehat{u}^{(i)}(x) - u^*(x)| < \epsilon \label{eq_4_3.6} \\
& \sup_{x \in \mathcal{X}} |\widehat{w}^{(i)}(x) - w^*(x)| < \epsilon \label{eq_4_3.7}.
\end{flalign} 
\end{theorem} 
\noindent \textbf{Proof.} By following the same proof procedures in Theorems 3 and 4 in  \cite{abu2005nearly}, the results \eqref{eq_4_3.5}-\eqref{eq_4_3.7} can be proven directly. Similar to \eqref{eq_4_3.6}, the result \eqref{eq_4_3.7} can also be proven.
\begin{remark} \label{remark7}
\emph{The proposed off-policy RL method is to learn the solution of the HJI equation \eqref{HJI} and the $ H_\infty $ control policy \eqref{idea_optimal_control}.  It follows from Theorem \ref{theorem_4.3} that the control policy $ \widehat{u}^{(i)} $ designed by the off-policy RL will uniformly converge to the $ H_\infty $ control policy $ u^* $. With the  $ H_\infty $ control policy, it is noted from Lemma \ref{lemma_1} that the closed-loop system \eqref{eq_sys} with $ w(t)\equiv 0 $ is locally asymptotically stable. Furthermore, it is observed from \eqref{L2_gain} that for the closed-loop system with disturbance $ w(t) \in L_2 [0,\infty) $, the output $z(t)$ is in $ L_2 [0,\infty) $ \cite{green1995linear}, i.e., the closed-loop system is (bounded-input bounded-output) stable.$ \square $ }
\end{remark}

\section{Off-policy reinforcement learning for linear $ H_\infty $ control} \label{Sec_5}
In this section, the developed NN-based off-policy RL method (i.e., Algorithm \ref{algorithm_2}) is simplified for linear $ H_\infty $ control design. Consider the linear system:
\begin{flalign}
\dot{x}(t) =& Ax(t) + B_2u(t) + B_1w(t) \label{eq_5.1}\\
z(t) =& Cx \label{eq_5.2}
\end{flalign}
where $A \in \mathbb{R}^{n\times n} $, $B_1 \in \mathbb{R}^{n\times q} $, $B_2 \in \mathbb{R}^{n\times m} $ and $C \in \mathbb{R}^{p\times n} $. Then, the HJI equation \eqref{HJI} of the linear system \eqref{eq_5.1} and \eqref{eq_5.2} results in an algebraic Riccati equation (ARE) \cite{green1995linear, wu2013simultaneous}: 
\begin{equation} \label{ARE}
A^TP + PA + Q + \gamma^{-2} PB_1B_1^TP - PB_2R^{-1}B_2^TP =0
\end{equation}
where $ Q = C^TC $. If ARE \eqref{ARE} has a stabilizing solution $ P \geqslant 0 $, the solution of the HJI equation \eqref{HJI} of the linear system \eqref{eq_5.1} and \eqref{eq_5.2} is $ V^*(x) = x^TPx $, and then the linear $ H_\infty $ control policy \eqref{idea_optimal_control}  is accordingly given by 
\begin{equation} \label{linear_H_policy}
u^*(x) = -R^{-1}B_2^TPx.
\end{equation}

Consequently, $ V^{(i)}(x) = x^TP^{(i)}x $, then the iterative equations \eqref{eq_control_update}-\eqref{LFE} in Algorithm \ref{algorithm_1} are respectively represented with
\begin{flalign}
u^{(i)} =& -R^{-1} B_2^T P^{(i)} x \label{eq_5.3} \\ 
w^{(i)} =& \gamma^{-2} B_1^T P^{(i)} x  \label{eq_5.4}
\end{flalign} 
\begin{equation} \label{eq_5.5}
\overline{A}_i^T P^{(i+1)} + P^{(i+1)} \overline{A}_i + \overline{Q}^{(i)} = 0 
\end{equation}
where $ \overline{A}_i \triangleq A + \gamma^{-2} B_1 B_1^T P^{(i)} - B_2 R^{-1} B_2^T P^{(i)} $ and $ \overline{Q}^{(i)} \triangleq Q -\gamma^{-2} P^{(i)} B_1 B_1^T P^{(i)} $ $+ P^{(i)} B_2 R^{-1}$ $ B_2^T P^{(i)} $.

Similar to the derivation of the off-policy RL method for nonlinear $ H_\infty $ control design in Section \ref{Sec_4}, rewrite the linear system \eqref{eq_5.1}  as
 \begin{equation}\label{eq_5.6}
\dot{x} = Ax + B_2u^{(i)} + B_1w^{(i)} + B_2[u - u^{(i)}] + B_1[ w-w^{(i)} ].
\end{equation}
Based on equations \eqref{eq_5.3}-\eqref{eq_5.6}, the equation \eqref{eq_4_3} is given by
\begin{flalign}\label{eq_5.7}
&\int_{t}^{t+\Delta t} x^T(\tau) P^{(i+1)} B_2 \left[ u(\tau) + R^{-1} B_2^T P^{(i)} x(\tau) \right] d \tau \nonumber \\
&\qquad + \int_{t}^{t+\Delta t} x^T(\tau) P^{(i+1)} B_1 \left[ w(\tau) - \gamma^{-2} B_1^T P^{(i)} x (\tau) \right] d \tau \nonumber \\
&\qquad + [x(t) - x(t+\Delta t)]^T P^{(i+1)} [x(t) - x(t+\Delta t)] \nonumber \\ 
&\quad = \int_{t}^{t+\Delta t}  x^T(\tau) \overline{Q}^{(i)} x(\tau) d\tau
\end{flalign} 
where $ P^{(i+1)} $ is a $ n\times n $ unknown matrix to be learned. For notation simplicity, define
\begin{flalign}
\rho_{\Delta x}(x(t)) \triangleq & x(t) - x(t+\Delta t) \nonumber \\
\rho_{xx} (x(t)) \triangleq & \int_{t}^{t+\Delta t} x(\tau) \otimes x(\tau)d \tau \nonumber \\
\rho_{ux} (x(t),u(t)) \triangleq & \int_{t}^{t+\Delta t} u(\tau) \otimes x(\tau)d \tau \nonumber \\
\rho_{wx} (x(t),w(t)) \triangleq & \int_{t}^{t+\Delta t} w(\tau) \otimes x(\tau)d \tau \nonumber
\end{flalign} 
where $ \otimes $ denotes Kronecker product. Each term of equation \eqref{eq_5.7}  can be written as:
\begin{flalign}
& \int_{t}^{t+\Delta t} x^T(\tau) P^{(i+1)} B_2 u(\tau) d \tau \nonumber \\
&\qquad = \rho^T_{ux} (x(t),u(t)) (B^T_2 \otimes I) vec( P^{(i+1)}) \nonumber \\
& \int_{t}^{t+\Delta t} x^T(\tau) P^{(i+1)} B_2 R^{-1} B_2^T P^{(i)} x(\tau) d \tau \nonumber \\
&\qquad = \rho^T_{xx} (x(t)) (P^{(i)} B_2 R^{-1}B^T_2 \otimes I) vec( P^{(i+1)}) \nonumber \\
& \int_{t}^{t+\Delta t} x^T(\tau) P^{(i+1)} B_1 w(\tau) d \tau \nonumber \\
&\qquad = \rho^T_{wx} (x(t),w(t)) (B^T_1 \otimes I) vec( P^{(i+1)}) \nonumber \\
& \gamma^{-2}  \int_{t}^{t+\Delta t} x^T(\tau) P^{(i+1)} B_1 B_1^T P^{(i)} x(\tau) d \tau \nonumber \\
&\qquad = \gamma^{-2}  \rho^T_{xx} (x(t)) (P^{(i)} B_1 B^T_1 \otimes I) vec( P^{(i+1)}) \nonumber \\
& [x(t) - x(t+\Delta t)]^T P^{(i+1)} [x(t) - x(t+\Delta t)] \nonumber \\
&\qquad = \rho^T_{\Delta x}(x(t)) vec( P^{(i+1)}) \nonumber \\
& \int_{t}^{t+\Delta t}  x^T(\tau) \overline{Q}^{(i)} x(\tau) d\tau  = \rho^T_{xx} (x(t)) vec( \overline{Q}^{(i)} ) \nonumber
\end{flalign} 
where $vec(P)$ denotes the vectorization of the matrix $ P $ formed by stacking the columns of $ P $ into a single column vector. Then, equation \eqref{eq_5.7} can be rewritten as
\begin{equation} \label{eq_5.8}
\overline{\rho}^{(i)}(x(t), u(t), w(t)) vec( P^{(i+1)}) = \pi^{(i)}(x(t))
\end{equation}
with
\begin{flalign}
& \overline{\rho}^{(i)}(x(t), u(t), w(t)) = \rho^T_{ux} (x(t),u(t)) (B^T_2 \otimes I) \nonumber \\
&\qquad + \rho^T_{wx} (x(t),w(t)) (B^T_1 \otimes I) +\rho^T_{\Delta x}(x(t)) \nonumber \\
&\qquad - \gamma^{-2} (P^{(i)} B_1 B^T_1 \otimes I) ] \nonumber \\
& \pi^{(i)}(x(t)) = \rho^T_{xx} (x(t)) vec( \overline{Q}^{(i)} ). \nonumber
\end{flalign}
It is noted that equation \eqref{eq_5.8} is equivalent to the equation \eqref{eq_4_13} with residual error $ \sigma ^{(i)} = 0 $. This is because no cost function approximation is required for linear systems. Then, by collecting sample set $ \mathcal {S}_M $ for computing $ \rho_{ux}, \rho_{wx}, \rho_{xx} $ and $ \rho_{\Delta x} $, a more simpler least-square scheme \eqref{eq_4_18} can be derived to obtain the unknown parameter vector $ vec( P^{(i+1)}) $ accordingly.
\begin{figure}
	\begin{minipage}[t]{0.5 \linewidth}
	\setcaptionwidth{1.6 in}
		\centering	\includegraphics[width=1.7in]{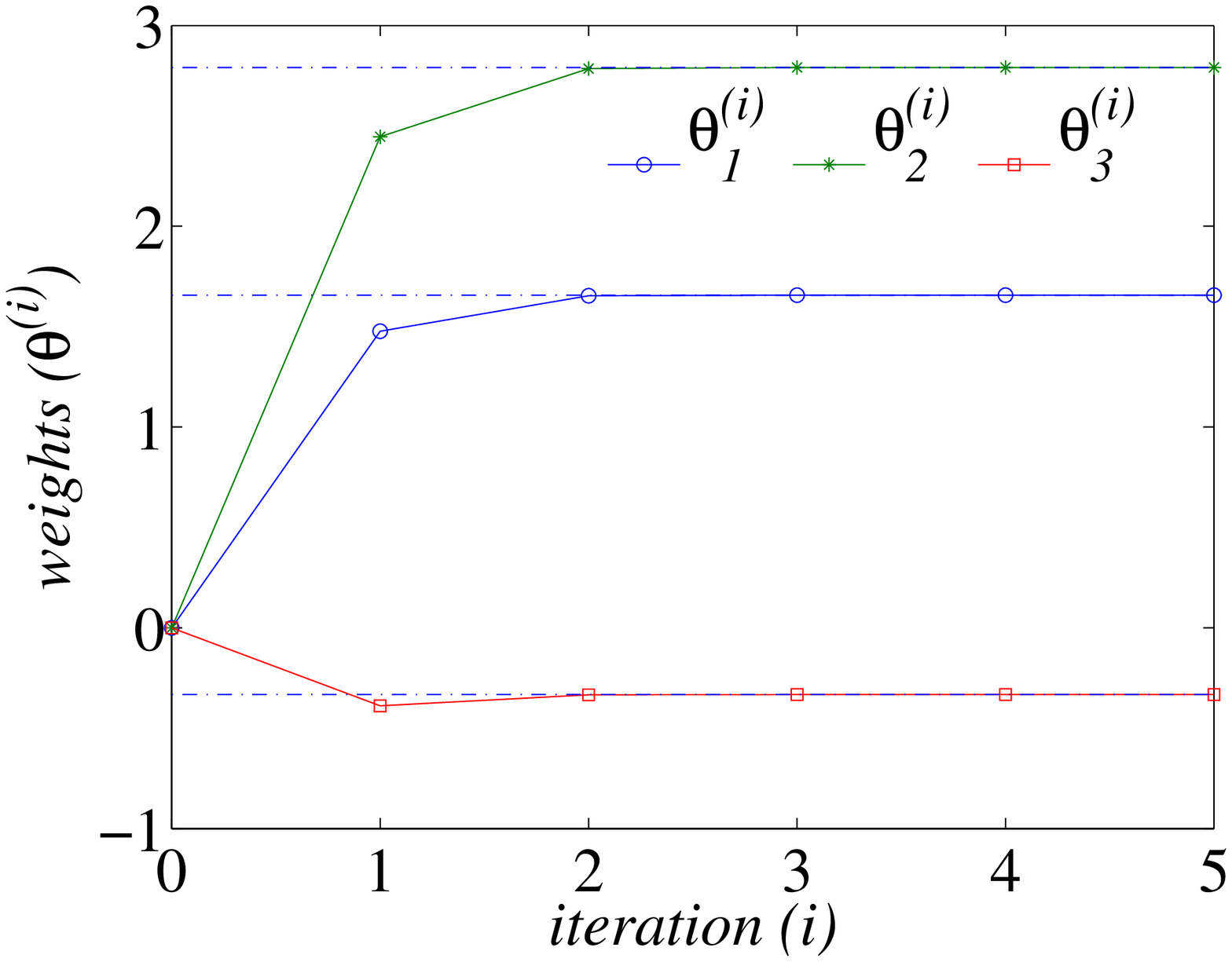}
	\caption{For the linear F16 aircraft plant, the critic NN weights $\theta_1^{(i)} \sim \theta_3^{(i)} $ at each iteration.}
		\label{fig1}
	\end{minipage}%
	\begin{minipage}[t]{0.5\linewidth}
	\setcaptionwidth{1.6 in}
		\centering	\includegraphics[width=1.7in]{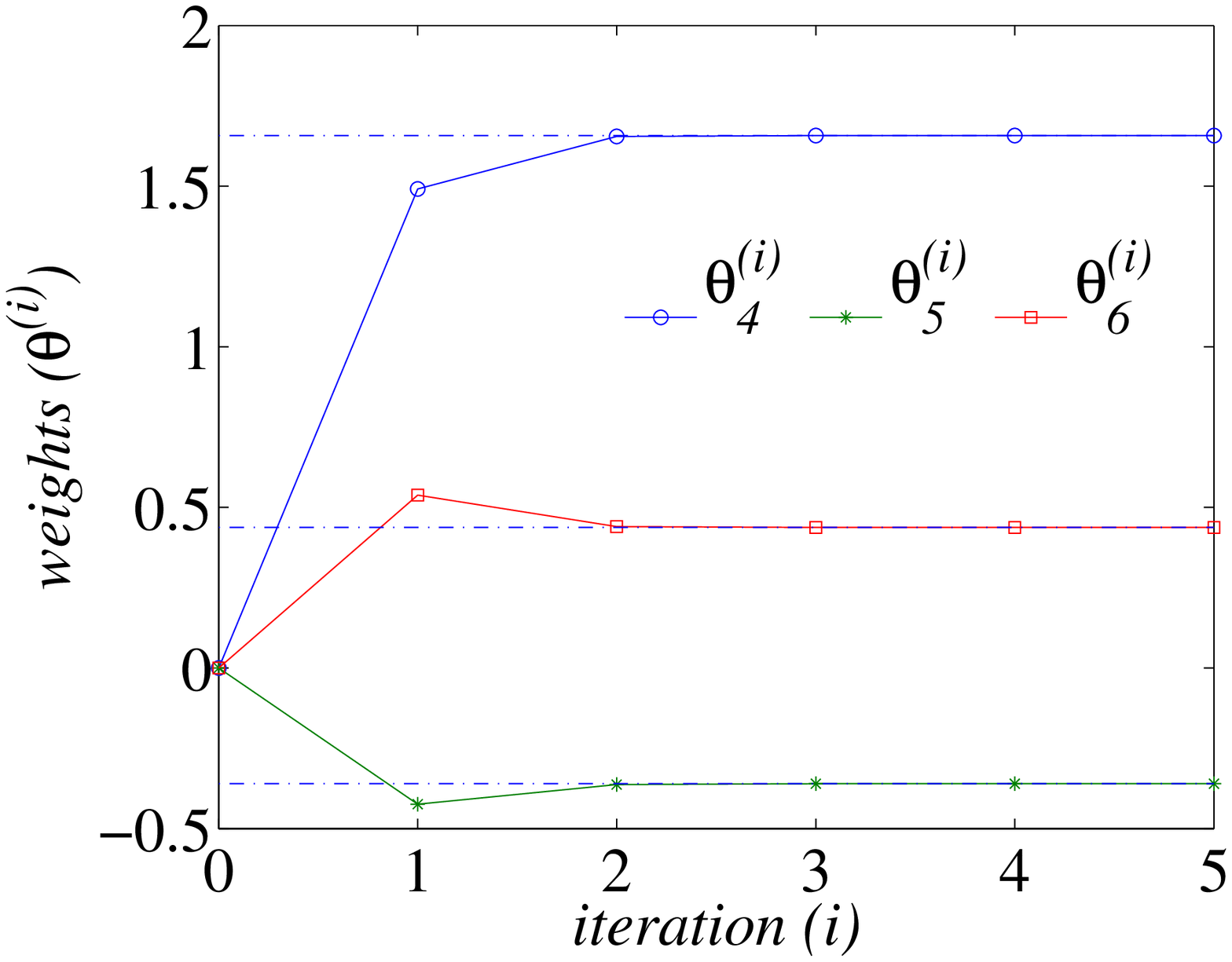}
		\caption{For the linear F16 aircraft plant, the critic NN  weights $\theta_4^{(i)} \sim \theta_6^{(i)} $ at each iteration.}
		\label{fig2}
	\end{minipage}
\end{figure}
\section{Simulation studies} \label{Sec_6}
In this section, the efficiency of the developed NN-based off-policy RL method is tested on a F16 aircraft plant. Then, it is applied to the rotational/translational actuator (RTAC) nonlinear benchmark problem.
\subsection{Efficiency test on linear F16 aircraft plant} \label{}
Consider a F16 aircraft plant that  used in \cite{stevens2003aircraft, vamvoudakis2012online, wu2012neural, wu2013simultaneous}, where the system dynamics is described by a linear continuous-time model:
\begin{flalign}
\dot{x} =& \left[ \begin{array}{ccc}
 		-1.01887 & 0.90506 & -0.00215 \\
 		0.82225 & -1.07741 & -0.17555 \\
 		0 & 0 & -1
	\end{array} \right] x\nonumber \\
& + 
\left[ \begin{array}{ccc}
 		0 \\
 		0 \\
 		1
	\end{array} \right] u+ 
\left[ \begin{array}{ccc}
 		1 \\
 		0 \\
 		0
	\end{array} \right] w \label{eq_6.1} \\ 
z = & x.\label{eq_6.2}
\end{flalign}
where the system state vector is $ x = [\alpha~q~\delta_e  ]^T $, $ \alpha $ denotes the angle of attack, $ q $ is the pitch rate and $ \delta_e $ is the elevator deflection angle. The control input $ u $ is the elevator actuator voltage and the disturbance $ w $ is wind gusts on angle of attack. Select $ R = 1 $ and $ \gamma = 5 $ for the $ L_2 $-gain performance \eqref{L2_gain}. Then, solve the associated ARE \eqref{ARE} with the MATLAB command CARE, we obtain
\begin{equation}
P=
 \left[ \begin{array}{ccc}
 		1.6573 & 1.3954 & -0.1661 \\
 		1.3954 & 1.6573 & -0.1804 \\
 		-0.1661 & -0.1804 & 0.4371
\end{array} \right]. \nonumber
\end{equation}
\begin{figure}
	\begin{minipage}[t]{0.5 \linewidth}
	\setcaptionwidth{1.6 in}
		\centering	\includegraphics[width=1.7in]{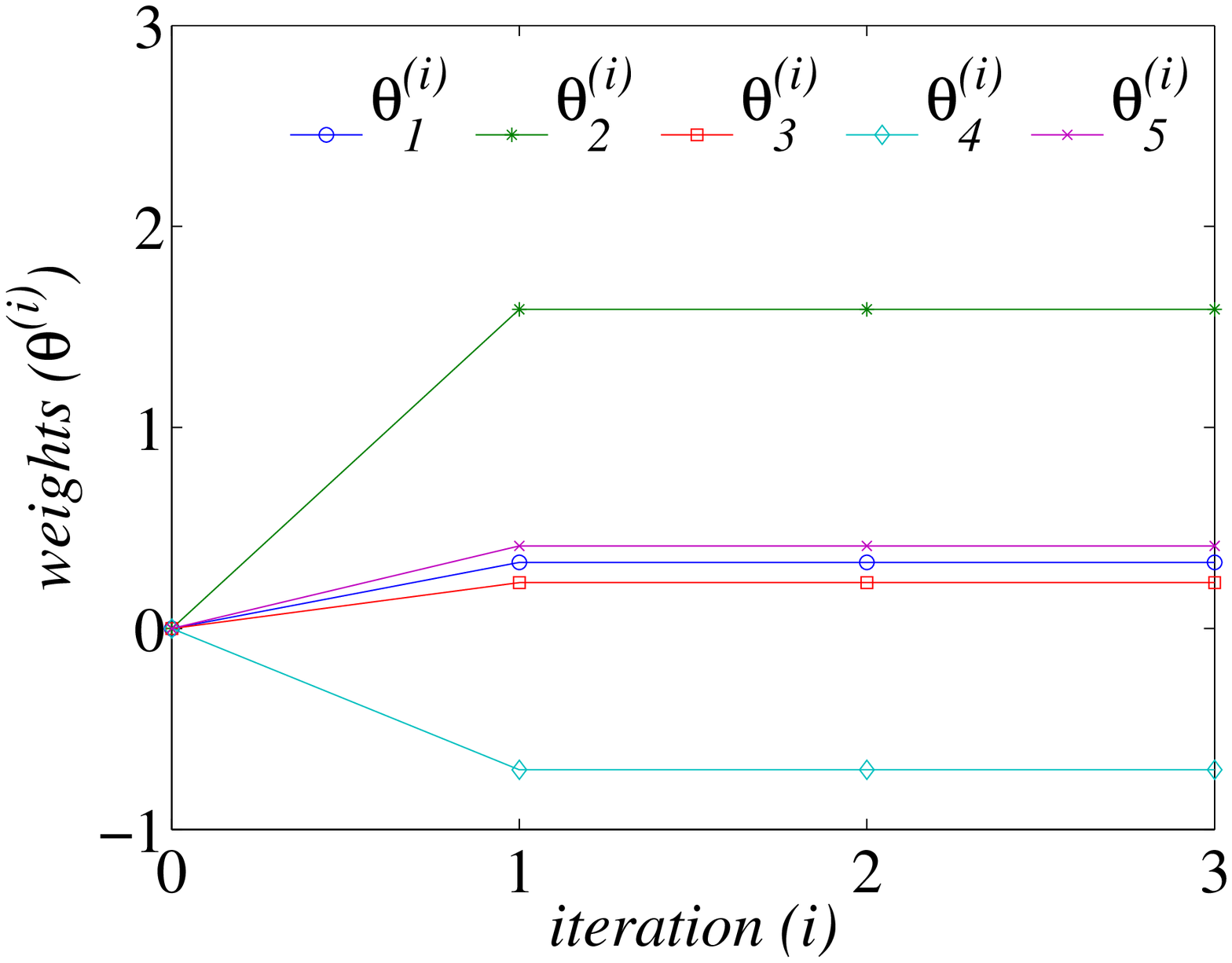}
		\caption{For the RTAC system, the critic NN  $\theta_1^{(i)} \sim \theta_5^{(i)} $ weights at each iteration. }
		\label{fig3}
	\end{minipage}%
	\begin{minipage}[t]{0.5\linewidth}
	\setcaptionwidth{1.6 in}
		\centering	\includegraphics[width=1.7in]{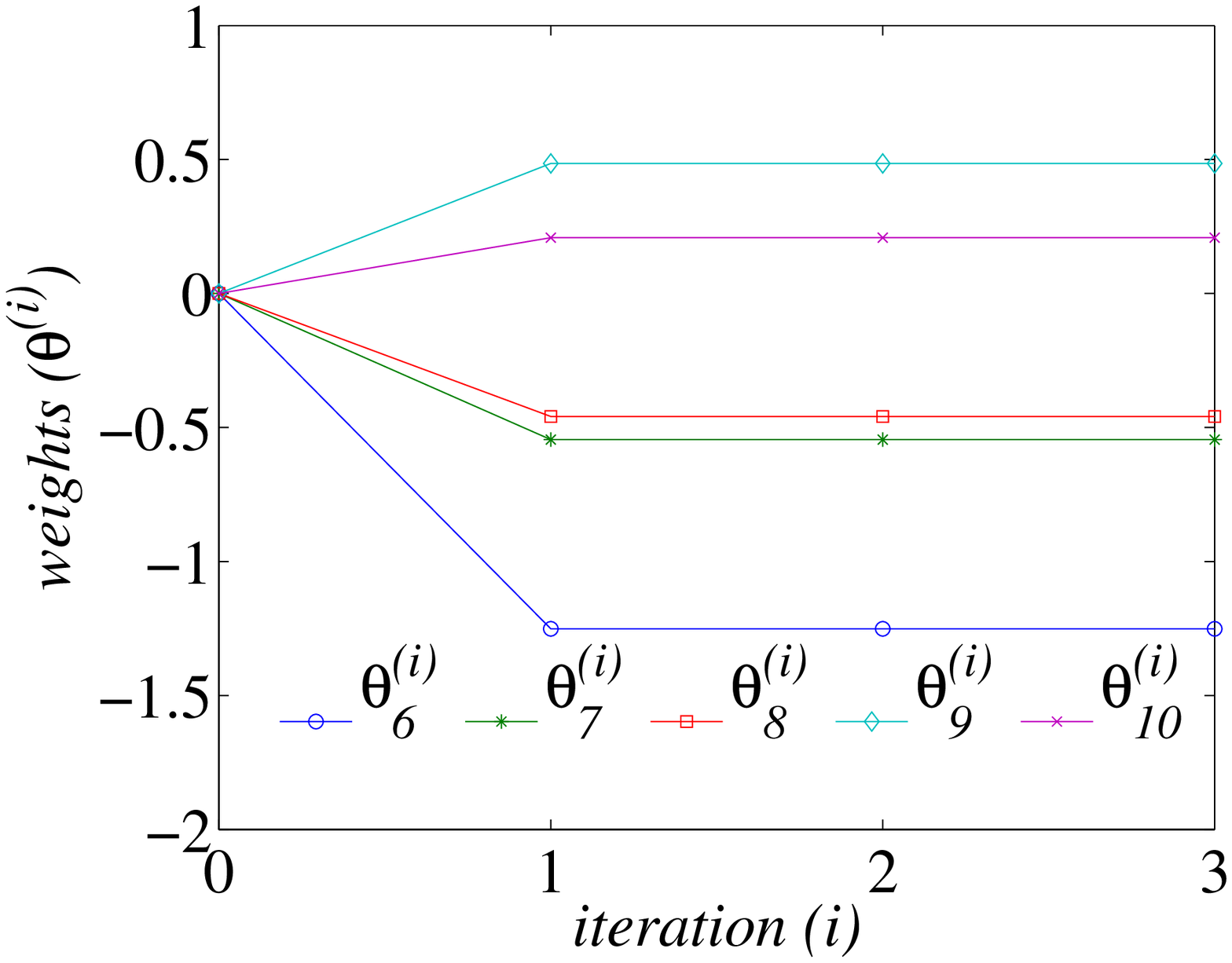}
		\caption{For the RTAC system, the critic NN $\theta_6^{(i)} \sim \theta_{10}^{(i)} $ weights at each iteration.} 
		\label{fig4}
	\end{minipage}
\end{figure} 
For linear systems, the  solution of the HJI equation is $ V^*(x) = x^TPx $, thus the complete activation function vector for critic NN is $\varphi(x) = \left[x_1^2~x_1x_2~x_1x_3~x_2^2~x_2x_3~x_3^2 \right]^T$ of size $ L=6 $. Then, the idea critic NN weight vector is $ \theta^* = [ p_{11}~2p_{12}~2p_{13}~p_{22}~2p_{23}~p_{33} ]^T $ $=  [1.6573~2.7908~$ $ -0.3322~ 1.6573~-0.3608~0.4371]^T$. Letting initial critic NN weight $ \theta_l^{(0)} = 0 (l = 1,...,6)  $, iterative stop criterion $ \xi = 10^{-7} $ and integral time interval $ \Delta t = 0.1s$, Algorithm \ref{algorithm_2} is applied to learn the solution of the ARE. 
To generate sample set   $ \mathcal {S}_M $, let sample size  $ M = 100 $ and generate random noise in interval [0,0.1] as input signals.
Figures \ref{fig1}-\ref{fig2} give the critic NN weight $\theta^{(i)} $ at each iteration, in which the dash lines represent idea values of $ \theta^* $. It is observed from the figures that the critic NN weight vector converges to the idea values of $ \theta^* $ at $ i = 5 $ iteration. Then, the efficiency of the developed off-policy RL method is verified.  

In addition, to test the influence of the parameter $ \Delta t $ to Algorithm \ref{algorithm_2}, we re-conduct simulation with different parameter cases: $ \Delta t = 0.2, 0.3, 0.4, 0.5s$, and the results show that the critic NN weight vector $\theta^{(i)} $ still converges to the idea values of $ \theta^* $ at $ i = 5 $ iteration for all cases. This implies that the developed off-policy RL algorithm \ref{algorithm_2} is insensitive to the parameter $ \Delta t $.

\subsection{Application to the rotational/translational actuator nonlinear benchmark problem}
 The RTAC nonlinear benchmark problem \cite{abu2008neurodynamic, tsiotras1998an, luo2013computationally} has been widely used to test the abilities of control methods. The dynamics of this nonlinear plant poses challenges because the rotational and translation motions are coupled. The RTAC system is given as follows:
\begin{flalign}
\dot{x} = &\left[ \begin{array} {*{3}{>{\displaystyle}c}}
 		x_2\\
 		\frac{-x_1 + \zeta x_4^2 \sin x_3}{1 - \zeta ^2 \cos^2 x_3} \\
 		x_4\\
 		\frac{\zeta \cos x_3 (x_1 - \zeta x_4^2 \sin x_3)}{1 - \zeta ^2 \cos^2 x_3}
	\end{array} \right] +
\left [ \begin{array} {*{3}{>{\displaystyle}c}}
 		0 \\
 		\frac{- \zeta \cos x_3}{1 - \zeta ^2 \cos^2 x_3} \\
 		0\\
 		\frac{1}{1 - \zeta ^2 \cos^2 x_3}
	\end{array} \right] u \nonumber \\
	&+ 
\left[ \begin{array} {*{3}{>{\displaystyle}c}}
 		0 \\
 		\frac{1}{1 - \zeta ^2 \cos^2 x_3} \\
 		0\\
 		\frac{- \zeta \cos x_3}{1 - \zeta ^2 \cos^2 x_3}
	\end{array} \right]w \label{eq_simulation_11} \\
z =& \sqrt{0.1} I x\label{eq_simulation_12}
\end{flalign}
where  $ \zeta =0.2 $. For the $ L_2 $-gain performance \eqref{L2_gain}, let $ R = 1 $ and $ \gamma = 6 $.
 \begin{figure}
	\begin{minipage}[t]{0.5 \linewidth}
	\setcaptionwidth{1.6 in}
		\centering	\includegraphics[width=1.7in]{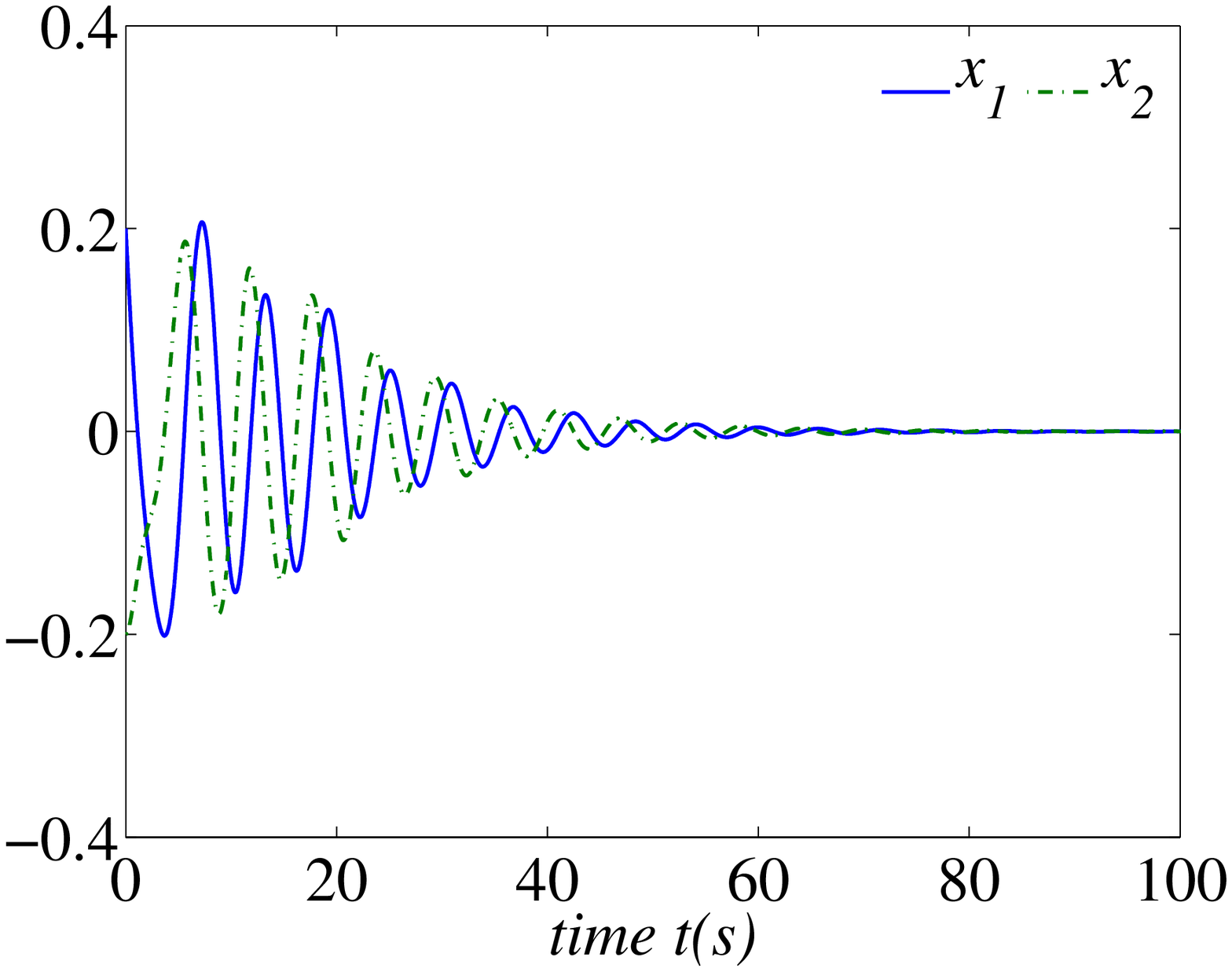}
		\caption{The state trajectories $x_1(t), x_2(t)$ of the closed-loop RTAC system. }
		\label{fig5}
	\end{minipage}%
	\begin{minipage}[t]{0.5\linewidth}
	\setcaptionwidth{1.6 in}
		\centering	\includegraphics[width=1.7in]{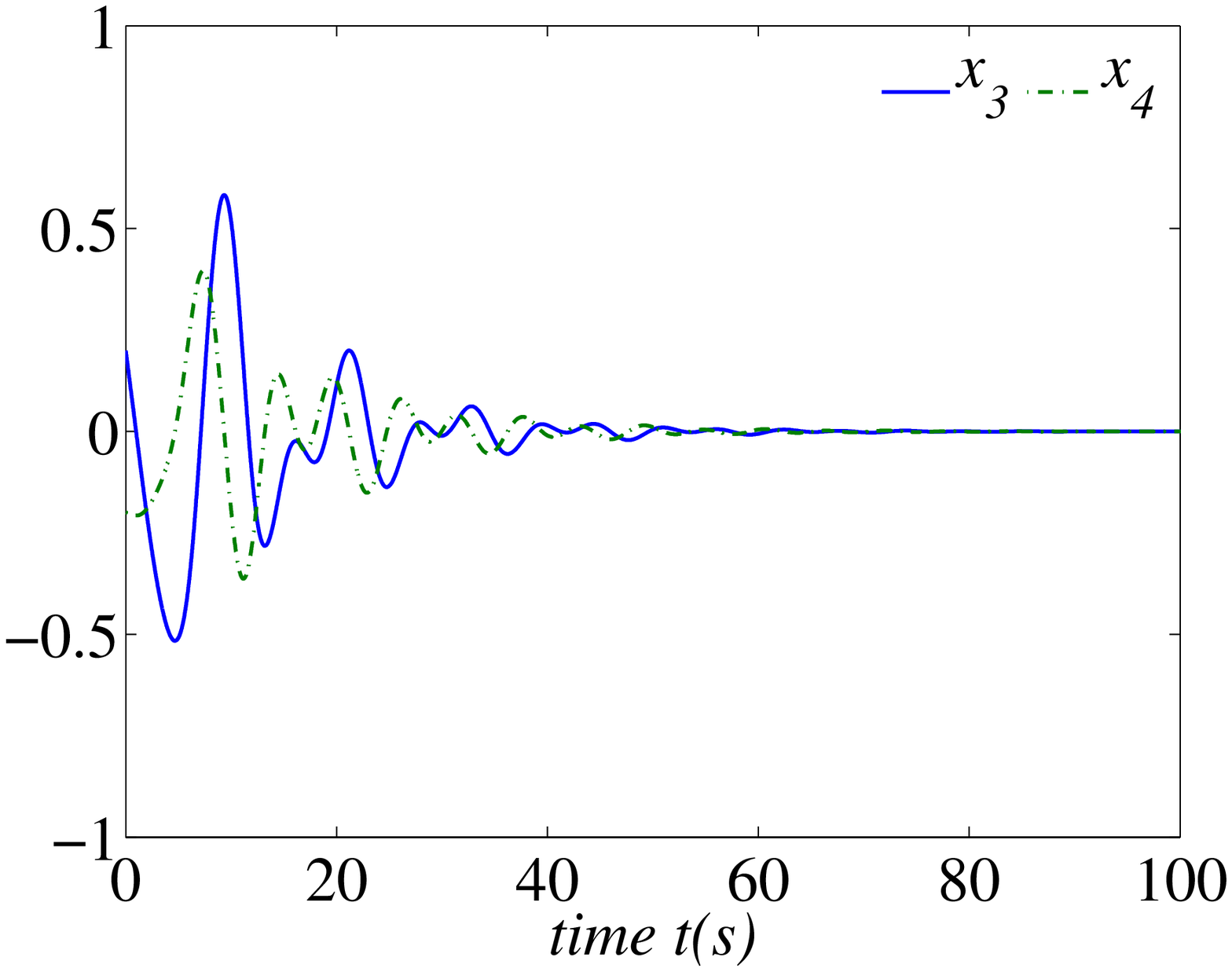}
		\caption{The state trajectories $x_3(t), x_4(t)$ of the closed-loop RTAC system.}
		\label{fig6}
	\end{minipage}
\end{figure} 
Then, the developed off-policy RL method is used to solve the nonlinear  $ H_\infty $ control problem of system \eqref{eq_simulation_11} and \eqref{eq_simulation_12}. Select the critic NN activation function vector as 
\begin{align}
\varphi(x) = [ x_1^2~~x_1x_2~~x_1x_3~~x_1x_4~~x_2^2~~x_2x_3~~x_2x_4~~x_3^2~~x_3x_4\nonumber\\ 
~~x_4^2~~x_1^3x_2~~x_1^3x_3~~x_1^3x_4~~x_1^2x_2^2~~x_1^2x_2x_3~~x_1^2x_2x_4~~x_1^2x_3^2\nonumber\\
~~x_1^2x_3x_4~~x_1^2x_4^2~~x_1x_2^3 ]^T \nonumber
\end{align}
of size $ L=20 $. With the initial critic NN weight $ \theta_l^{(0)} = 0 (l = 1,...,20)  $, iterative stop criterion $ \xi = 10^{-7} $ and integral time interval $ \Delta t = 0.033s$, Algorithm \ref{algorithm_2} is applied to learn the solution of the HJI equation. 
To generate sample set   $ \mathcal {S}_M $, let sample size  $ M = 300 $ and generate random noise in interval [0,0.5] as input signals.
It is found that the critic NN weight vector converges fast to 
\begin{align}
\theta^{(3)} = [0.3285~1.5877~ 0.2288~-0.7028~0.4101~-1.2514 \nonumber \\
~-0.5448~-0.4595~0.4852~0.2078~-1.3857~1.7518\nonumber \\
~1.1000~0.5820~0.1950~-0.0978~-1.0295~-0.2773\nonumber \\
~-0.2169~0.2463]^T \nonumber
\end{align}
at $ i = 3 $ iteration. Figures \ref{fig3}-\ref{fig4} show first 10 critic NN weights (i.e., $\theta_1^{(i)} \sim \theta_{10}^{(i)} $) at each iteration. With the convergent critic NN weight vector $ \theta^{(3)} $, the $ H_\infty $ control policy can be computed with \eqref{eq_4_10}. Under the disturbance signal $ w(t) = 0.2r_1(t)e^{-0.2t}cos(t), (r_1(t) \in [0,1]$ is a random number), closed-loop simulation is conducted with the $ H_\infty $ control policy. Figures \ref{fig5}-\ref{fig7} give the trajectories of state and control policy. To show the relationship between $ L_2 $-gain and time, define the following ratio of disturbance attenuation as
\begin{equation} {\displaystyle}
r_d(t) = \left( \frac{\int_{0}^{t} {\left( \Vert z(\tau) \Vert ^2 + \Vert u(\tau) \Vert _R^2 \right)} d \tau}{\int_{0}^{t} {\Vert w(\tau) \Vert ^2} d \tau} \right) ^{\frac{1}{2}}. \nonumber
\end{equation}
Figure \ref{fig8} shows the curve of $ r_d(t) $ , where it converges to 3.7024$(<\gamma = 6)$ as time increases, which implies that the designed $ H_\infty $ control law can achieve a prescribed $ L_2 $-gain performance level $ \gamma $ for the closed-loop system.
\begin{figure}
	\begin{minipage}[t]{0.5 \linewidth}
	\setcaptionwidth{1.6 in}
		\centering	\includegraphics[width=1.7in]{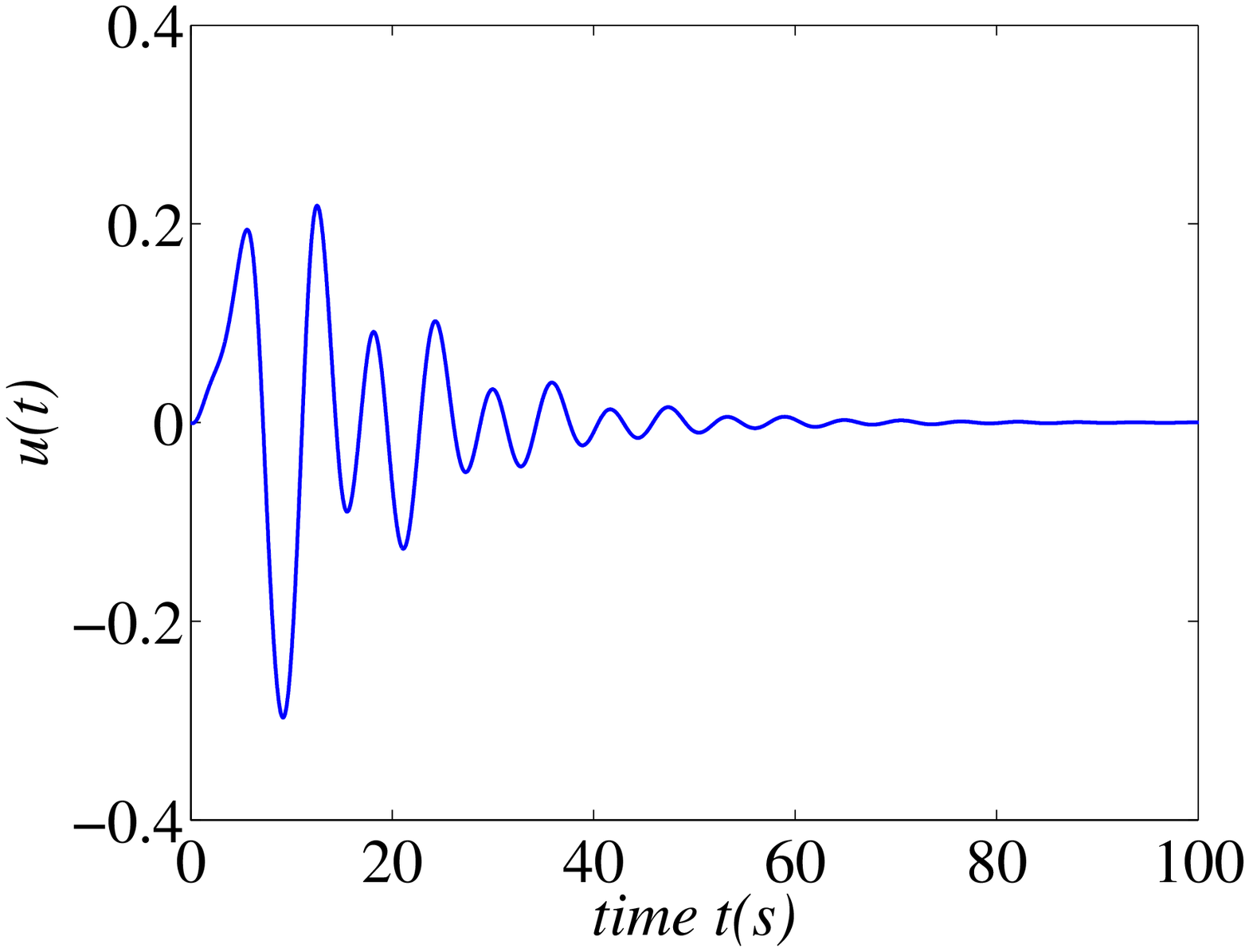}
		\caption{The control trajectory $u(t)$ of the closed-loop RTAC system.}
		\label{fig7}
	\end{minipage}%
	\begin{minipage}[t]{0.5\linewidth}
	\setcaptionwidth{1.6 in}
		\centering	\includegraphics[width=1.7in]{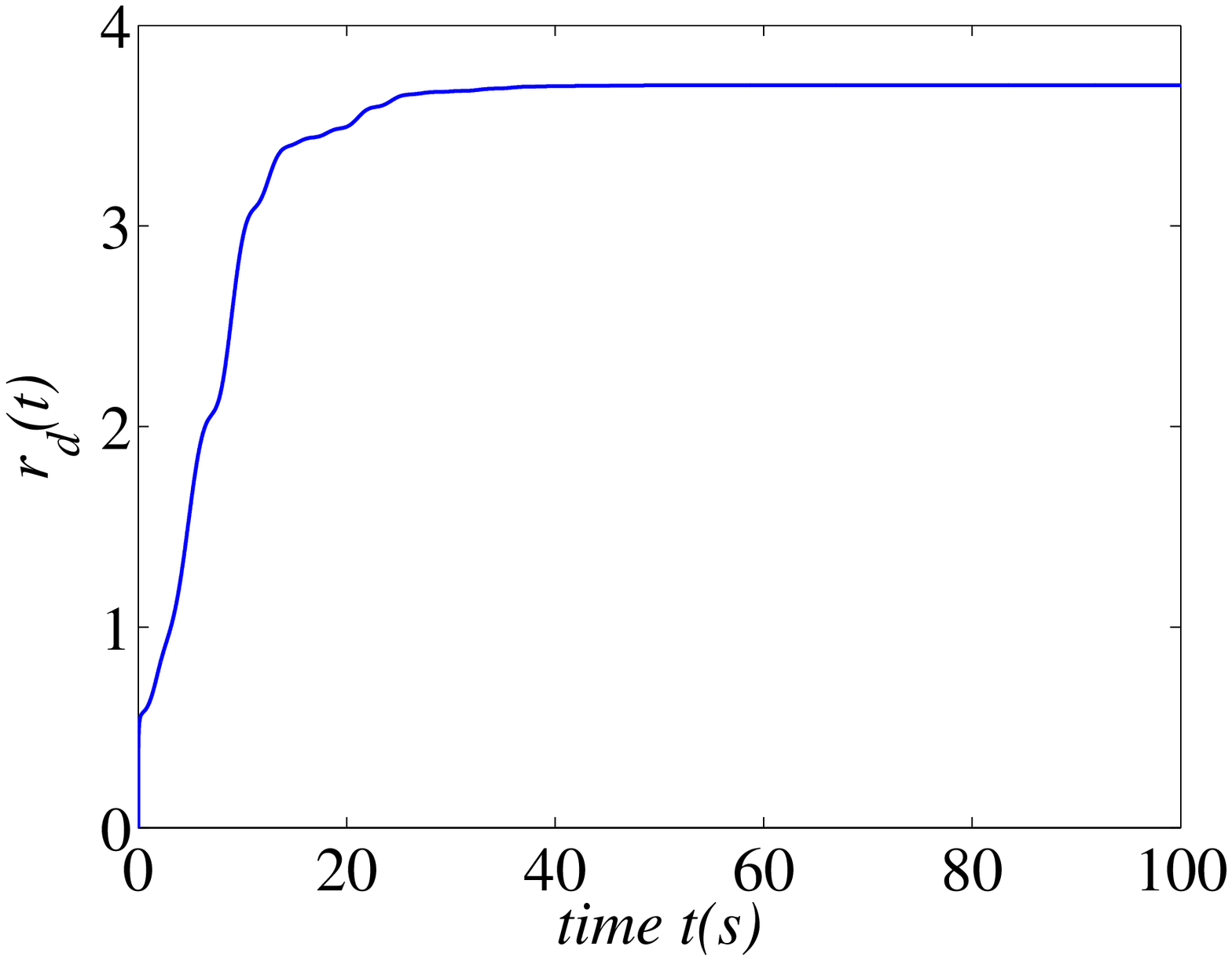}
		\caption{The curve $r_d(t)$ of the closed-loop RTAC system.}
		\label{fig8}
	\end{minipage}
\end{figure} 
\section{Conclusions} \label{Sec_7}
A NN-based off-policy RL method has been developed to solve the $H_\infty$ control problem of continuous-time systems with unknown internal system model. Based on the model-based SPUA, an off-policy RL method is derived, which can learn the solution of HJI equation from the system data generated by arbitrary control and disturbance signals. The implementation of the off-policy RL method is based on an actor-critic structure, where only one NN is required for approximating the cost function, and then a least-square scheme is derived for NN weights update. The effectiveness of the proposed NN-based off-policy RL method is tested on a linear F16 aircraft plant and a nonlinear RTAC problem.


%

%


\ifCLASSOPTIONcaptionsoff
  \newpage
\fi



%


%
\begin{IEEEbiography}[{\includegraphics[width=1in,height=1.25in,clip,keepaspectratio]{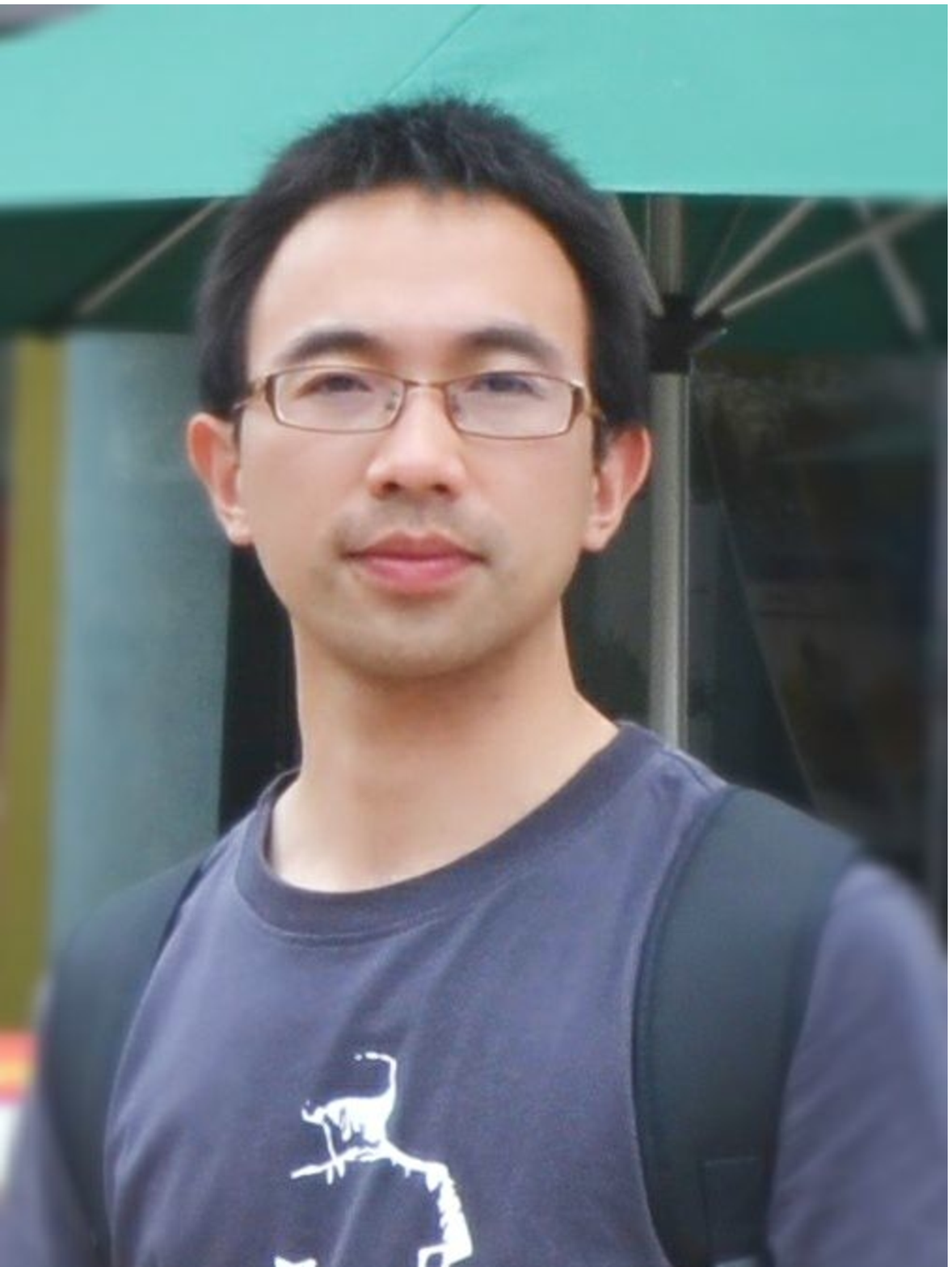}}]{Biao Luo}
received the B.E. degree in Measuring and Control Technology and Instrumentations and the M. E. degree in Control Theory and Control Engineering from Xiangtan University, Xiangtan, China, in 2006 and 2009, respectively. He is now working for the Ph. D. degree in Control Science and Engineering with Beihang University (Beijing University of Aeronautics and Astronautics), Beijing, China.

From February 2013 to August 2013, he was a Research Assistant with the Department of System Engineering and Engineering Management (SEEM), City University of Hong Kong, Kowloon, Hong Kong. From September 2013 to December 2013, he was a Research Assistant with Department of Mathematics and Science, Texas A\&M University at Qatar, Doha, Qatar. His current research interests include distributed parameter systems, optimal control, data-based control, fuzzy/neural modeling and control, hypersonic entry/reentry guidance, learning and control from big data, reinforcement learning, approximate dynamic programming, and evolutionary computation.

Mr. Luo was a recipient of the Excellent Master Dissertation Award of Hunan Province in 2011.
\end{IEEEbiography}

\begin{IEEEbiography}[{\includegraphics[width=1in,height=1.25in,clip,keepaspectratio]{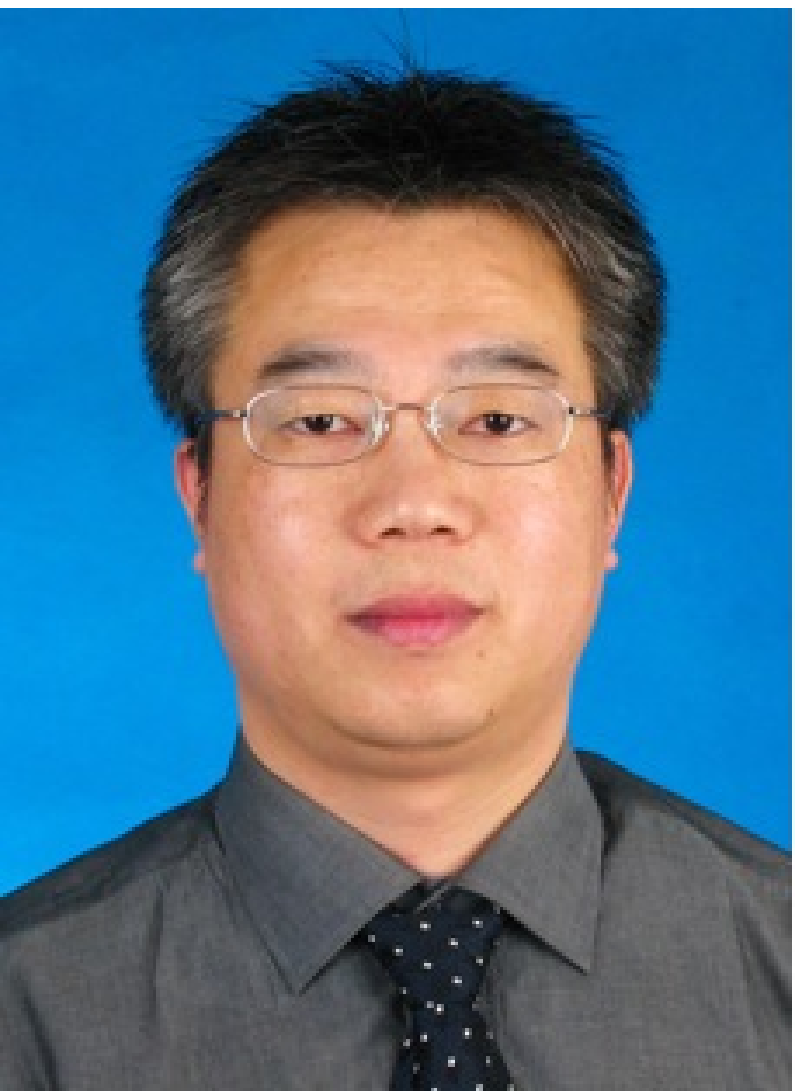}}]{Huai-Ning Wu}
was born in Anhui, China, on November 15, 1972. He received the B.E. degree in automation from Shandong Institute of Building Materials Industry, Jinan, China and the Ph.D. degree in control theory and control engineering from Xi’an Jiaotong University, Xi’an, China, in 1992 and 1997, respectively.

From August 1997 to July 1999, he was a Postdoctoral Researcher in the Department of Electronic Engineering at Beijing Institute of Technology, Beijing, China. In August 1999, he joined the School of Automation Science and Electrical Engineering, Beihang University (formerly Beijing University of Aeronautics and Astronautics), Beijing. From December 2005 to May 2006, he was a Senior Research Associate with the Department of Manufacturing Engineering and Engineering Management (MEEM), City University of Hong Kong, Kowloon, Hong Kong. From October to December during 2006-2008 and from July to August in 2010, he was a Research Fellow with the Department of MEEM, City University of Hong Kong. From July to August in 2011 and 2013, he was a Research Fellow with the Department of Systems Engineering and Engineering Management, City University of Hong Kong. He is currently a Professor with Beihang University. His current research interests include robust control, fault-tolerant control, distributed parameter systems, and fuzzy/neural modeling and control.

Dr. Wu serves as Associate Editor of the IEEE Transactions on Systems, Man \& Cybernetics: Systems. He is a member of the Committee of Technical Process Failure Diagnosis and Safety, Chinese Association of Automation.
\end{IEEEbiography}

\begin{IEEEbiography}[{\includegraphics[width=1in,height=1.25in,clip,keepaspectratio]{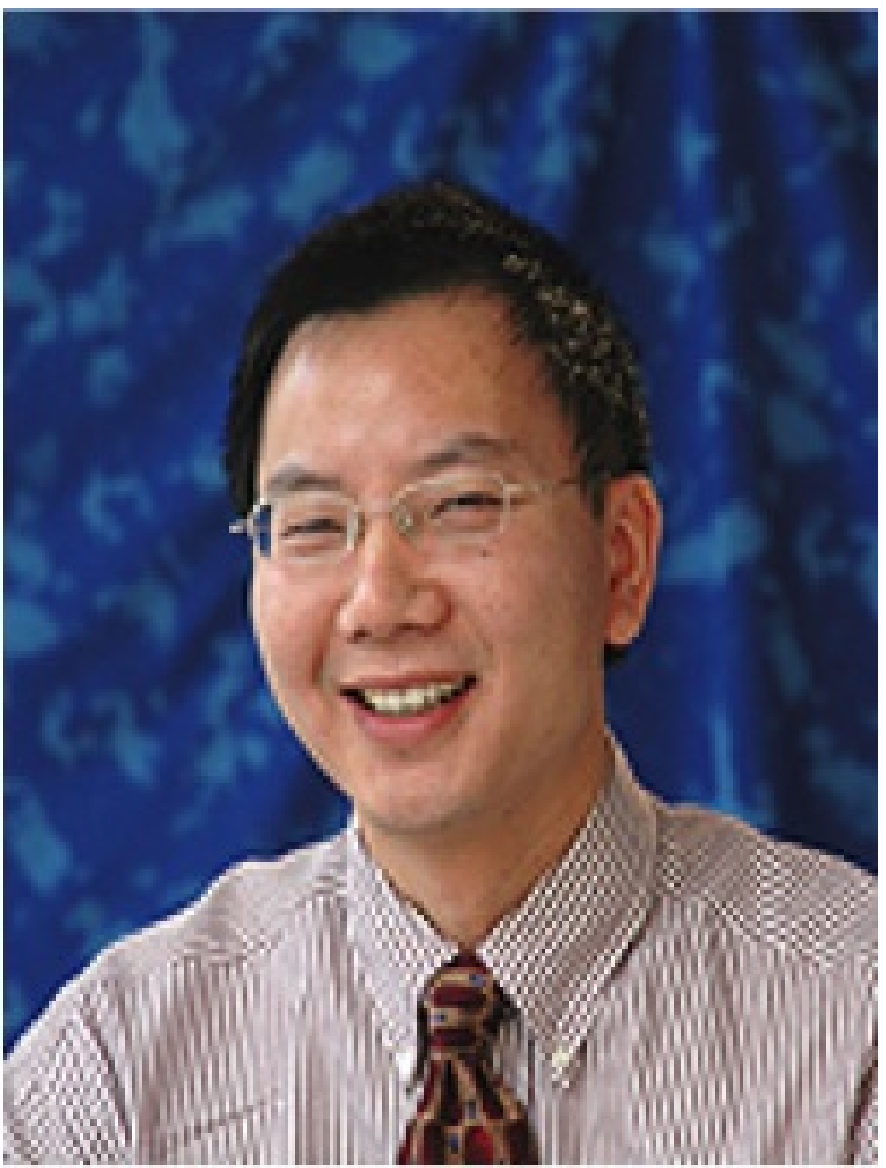}}]{Tingwen Huang}
is a Professor at Texas A\& M University at Qatar. He received his B.S. degree from Southwest Normal University (now Southwest University), China, 1990, his M.S. degree from Sichuan University, China, 1993, and his Ph.D. degree from Texas A\& M University, College Station, Texas, 2002. After graduated from Texas A\&M University, he worked as a Visiting Assistant Professor there. Then he joined Texas A\& M University at Qatar (TAMUQ) as an Assistant Professor in August 2003, then he was promoted to Professor in 2013. Dr. Huang’s focus areas for research interests include neural networks, chaotic dynamical systems, complex networks, optimization and control. He has authored and co-authored more than 100 refereed journal papers.
\end{IEEEbiography}




\end{document}